\documentclass{emulateapj}
\usepackage[breaklinks,colorlinks,urlcolor=blue,citecolor=blue,linkcolor=blue]{hyperref}
\usepackage{mathrsfs}
\usepackage{amsmath}
\usepackage{graphicx}

\shortauthors{Metzger et al.}

\shorttitle{Connecting FRB 121102 to Magnetar Birth}

\def\frb{\rm FRB\,121102}

\newcommand\degd{\ifmmode^{\circ}\!\!\!.\,\else$^{\circ}\!\!\!.\,$\fi}

\newcommand{\be}{\begin{equation}}
\newcommand{\ee}{\end{equation}}

\begin{document}

\title{Millisecond Magnetar Birth Connects FRB 121102 to Superluminous Supernovae and Long Duration Gamma-ray Bursts}

\author{
Brian D.~Metzger\altaffilmark{1},
Edo Berger\altaffilmark{2},
\& Ben Margalit\altaffilmark{1}
}

\altaffiltext{1}{Columbia Astrophysics Laboratory, Pupin Hall, New
  York, NY, 10027, USA; bmetzger@phys.columbia.edu}

\altaffiltext{2}{Harvard-Smithsonian Center for Astrophysics, 60
  Garden Street, Cambridge, MA 02138}

\begin{abstract}
Sub-arcsecond localization of the repeating fast radio burst FRB\,121102 revealed its coincidence with a dwarf host galaxy and a steady (`quiescent') non-thermal radio source.  We show that the properties of the host galaxy are consistent with those of long-duration gamma-ray bursts (LGRB) and hydrogen-poor superluminous supernovae (SLSNe-I).  Both LGRBs and SLSNe-I were previously hypothesized to be powered by the electromagnetic spin-down of newly-formed, strongly-magnetized neutron stars with millisecond birth rotation periods (`millisecond magnetars').  This motivates considering a scenario whereby the repeated bursts from \frb~originate from a young magnetar remnant embedded within a young hydrogen-poor supernova remnant.  Requirements on the dispersion measure and GHz free-free optical depth through the expanding supernova ejecta (accounting for photo-ionization by the rotationally-powered magnetar nebula), energetic constraints on the bursts, and constraints on the size of the quiescent source all point to an age of a few decades to a century.  The quiescent radio source can be attributed to synchrotron emission from the shock interaction between the fast outer layer of the supernova ejecta with the surrounding wind of the progenitor star, or from deeper within the magnetar wind nebula as outlined in Metzger et al.~(2014).  Alternatively, the radio emission could be an orphan afterglow from an initially off-axis LGRB jet, though this might require the source to be too young.  The young age of the source can be tested by searching for a time derivative of the dispersion measure and the predicted fading of the quiescent radio source.  We propose future tests of the SLSNe-I/LGRB/FRB connection, such as searches for FRBs from nearby SLSNe-I/LGRBs on timescales of decades after their explosions.   
 \end{abstract}

\keywords{stars:neutron$-$supernovae:general, galaxies: active – intergalactic medium – radio continuum: general – scattering}

\section{Introduction}

Fast radio bursts (FRB) are pulses of coherent GHz radio emission with durations of milliseconds or less and dispersion measures DM $\sim 500-1200$ pc cm$^{-3}$ much larger than expected for propagation through the Galaxy or its halo (\citealt{Lorimer+07,Keane+12,Thornton+13,Spitler+14,Ravi+15,Petroff+16,Champion+16}; see \citealt{Katz16} for a review).  Until recently, little was known about the sources of FRBs, even whether they were Galactic or extragalactic in origin.  This uncertainty resulted largely because the bursts were discovered primarily by single radio dishes with poor angular resolution.  Follow-up searches conducted in the error box of FRB 150418 led to a claimed detection of a host galaxy and a coincident radio counterpart \citep{Keane+16}; however, this association was disputed by \citet{Williams&Berger16}, who showed that the coincident source was an active galactic nucleus (AGN) with a significant probability of being located by chance within the FRB sky error region (see also \citealt{Johnston+17}).

Although most FRBs are detected as single bursts, several were discovered to repeat with consistent DM and sky localization (\citealt{Spitler+14,Spitler+16,Scholz+16}).  Using fast-dump interferometry with the Karl G.~Jansky Very Large Array (VLA) \citep{Law+15},  FRB\,121102 was recently localized to $\approx$0.1 arcsecond precision (\citealt{Chatterjee+17}).  Optical imaging and spectroscopy by \citet{Tendulkar+17} identify an extended source coincident with the burst displaying prominent Balmer and [OIII] lines at a redshift of $z = 0.19273$, indicating a dwarf galaxy of estimated stellar mass $M_{\star} \approx  4-7\times 10^{7}M_{\odot}$ at a luminosity distance of $D = 972$ Mpc $\simeq 3\times 10^{27}$ cm.  Based on H$\alpha$ flux of the galaxy, \citet{Tendulkar+17}  infer a total star formation rate of $0.4 M_{\odot}$ yr$^{-1}$, which they use to estimate a maximum DM through the plane of the galaxy of 324 pc cm$^{-3}$.  This is much higher than the residual (host + local) dispersion measure DM$_{\rm host+local} \approx  55-225$ pc cm$^{-3}$ (\citealt{Tendulkar+17}) when contributions from the Galaxy, Galactic halo and intergalactic medium as subtracted off the measured DM.  

In addition to localizing \frb, \citet{Chatterjee+17} identify a coincident continuum radio source with a relatively flat spectrum $F_{\nu} \approx \nu^{\beta}$ in the frequency range $\nu = 1.6-22$ GHz with $\beta \approx -0.2$ and a 1.7(5 GHz) luminosity of $\nu L_{\nu} \approx 3(7)\times 10^{38}$ erg s$^{-1}$, though the spectrum steepens at the highest frequencies $\nu \gtrsim 10$ GHz. \citet{Marcote+17} use VLBI observations to show that the quiescent radio source is co-located with the burst to $< 40$ pc (\citealt{Marcote+17}) and is constrained to a projected size at 5 GHz of $\lesssim 0.7$ pc.  Both radio sources are offset by $\sim 0.5-1$ kpc from the light centroid of the host galaxy \citep{Tendulkar+17}, potentially disfavoring an AGN or galactic nuclear origin (though the morphology of dwarf galaxies are highly irregular and hence their nuclei are not easy to locate).  {\it Chandra} observations place an upper limit of $L_{X} \lesssim 5\times 10^{41}$ erg s$^{-1}$ on the luminosity of spatially coincident X-ray emission, also in tension with attributing the compact radio source to an AGN \citep{Chatterjee+17}.

The repetition and energetics of the bursts from \frb~led \citet{Marcote+17} and \citet{Tendulkar+17} to argue a possible origin associated with a young neutron star or magnetar \citep{Kulkarni+15,Katz16,Cordes&Wasserman16,Lyutikov+16,Popov&Pshirkov16,Yang+16}, embedded within the ejecta shell of a young supernova (SN) remnant (\citealt{Connor+16,Piro16,Murase+16}).  In this paper we present further evidence in favor of an association between \frb~and the birth of a young magnetar.  

In $\S\ref{sec:host}$ we show quantitatively that the host galaxy of \frb~is consistent with those of hydrogen-poor superluminous SNe (SLSNe-I; \citealt{Quimby+11,GalYam12}) and long-duration gamma-ray bursts (LGRBs), a possible connection already pointed out by \citet{Tendulkar+17} and \citet{Marcote+17}.  In $\S\ref{sec:magnetar}$ we review the birth of a millisecond magnetars as proposed central engines of both long LGRBs (\citealt{Usov92,Wheeler+00,Thompson+04,Metzger+07,Metzger+11}) and SLSNe-I (\citealt{Kasen&Bildsten10,Woosley10}).  If such a magnetar is also capable of producing all of the observed features of \frb, then this places stringent constraints on the age of the system.  In $\S\ref{sec:ejecta}$ we show the expanding oxygen-rich SN ejecta will become transparent to GHz emission on a timescale as short as a decade after the explosion, while dispersion measure constraints are tighter and might suggest an older age approaching a century, depending on the fraction of the ejecta ionized by the proto-magnetar wind.  In $\S\ref{sec:radio}$ we describe possible sources for the quiescent radio counterpart within this picture on a timescale of decades after the explosion.  These include emission escaping directly through the ejecta shell from a nascent rotationally-powered ``magnetar wind nebula"; shock interaction between the fastest parts of the magnetar-boosted supernova ejecta with the surrounding stellar progenitor wind; or an orphan radio afterglow from an initially off-axis LGRB.  In \S\ref{sec:discussion} we discuss our results and expand on possible predictions of a connection between SLSNe-I/LGRBs and FRBs.  We briefly summarize our conclusions in $\S\ref{sec:conclusions}$.

\begin{figure*}
\includegraphics[width=.5\textwidth]{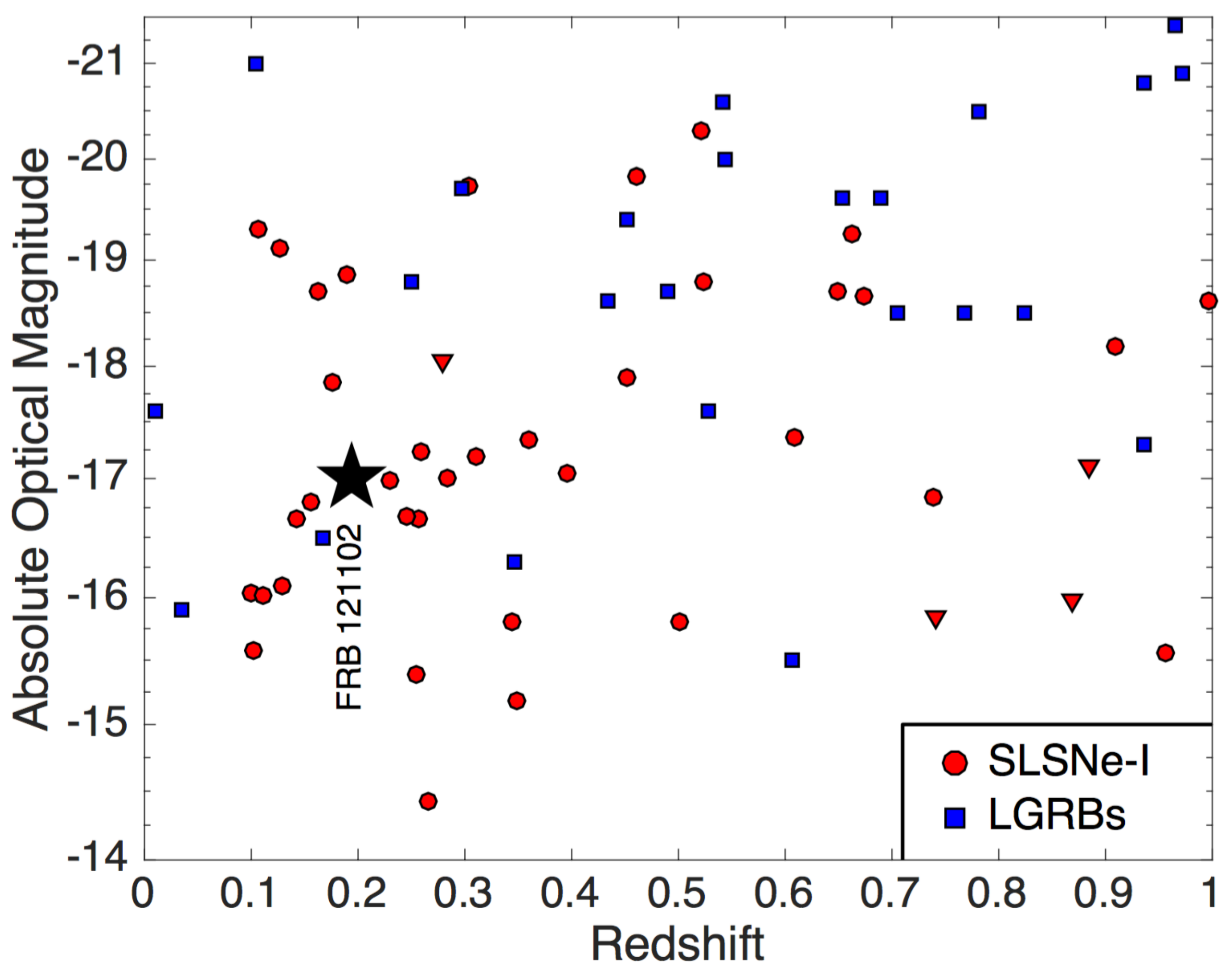}
\includegraphics[width=.5\textwidth]{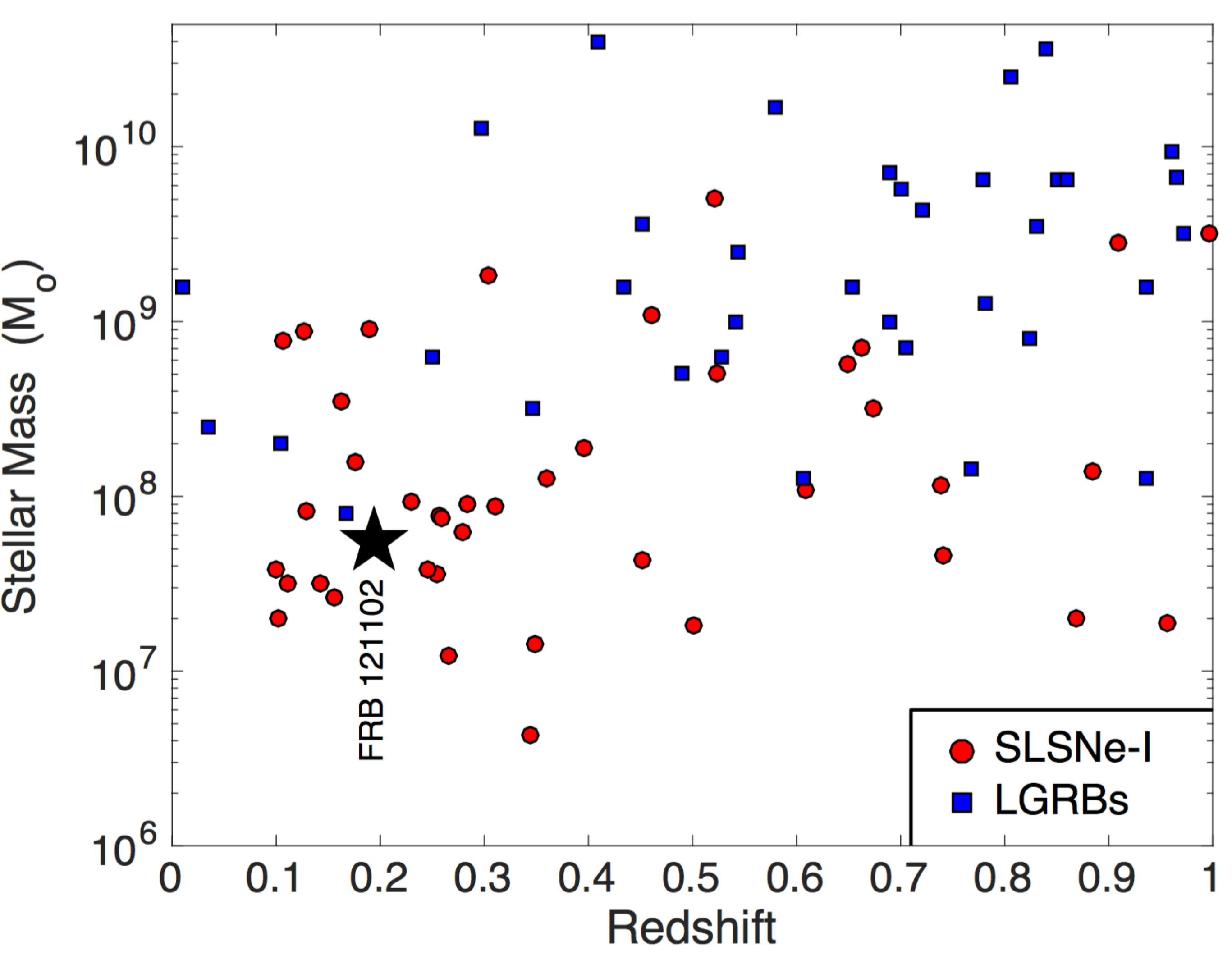}
\includegraphics[width=.5\textwidth]{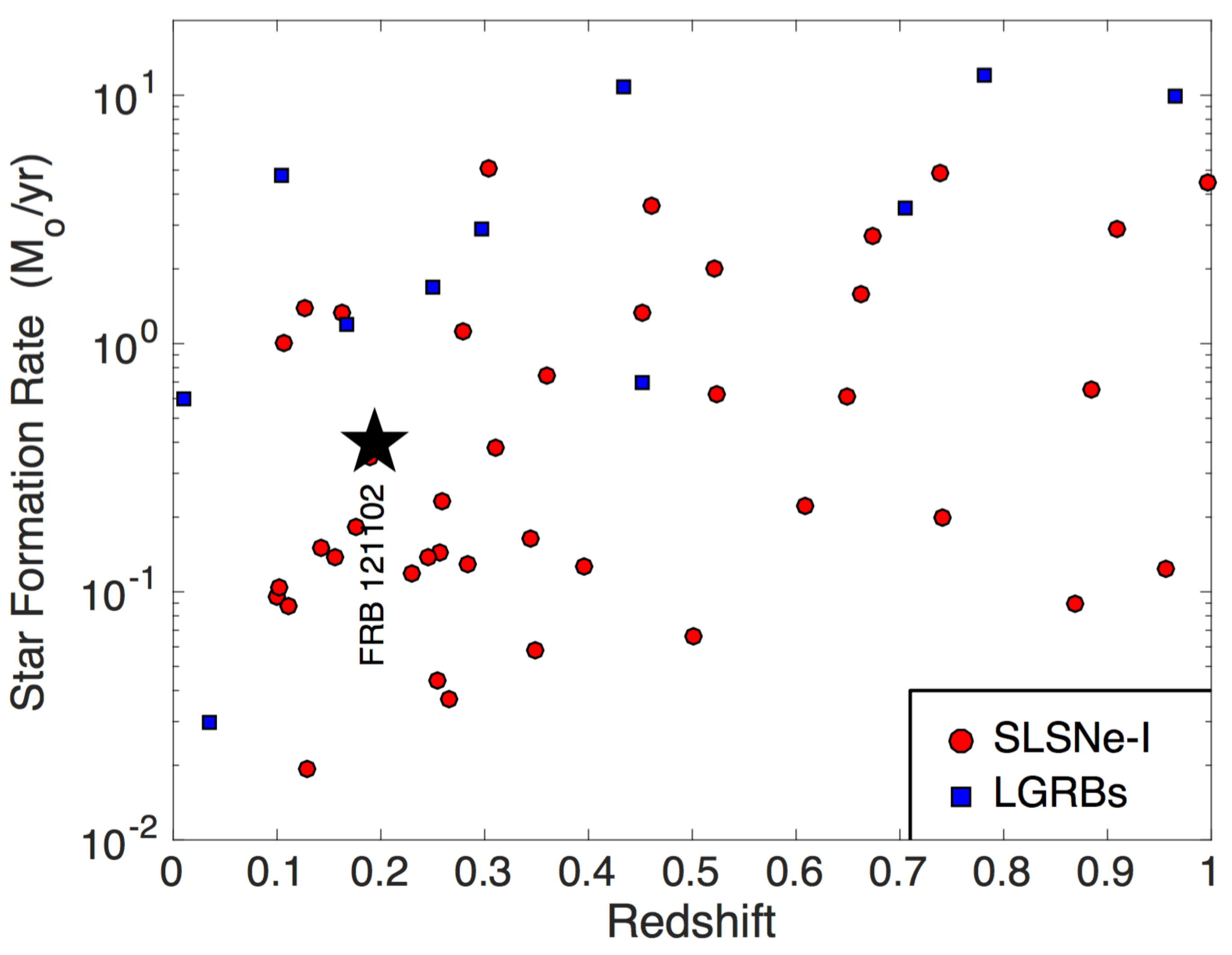}
\includegraphics[width=.5\textwidth]{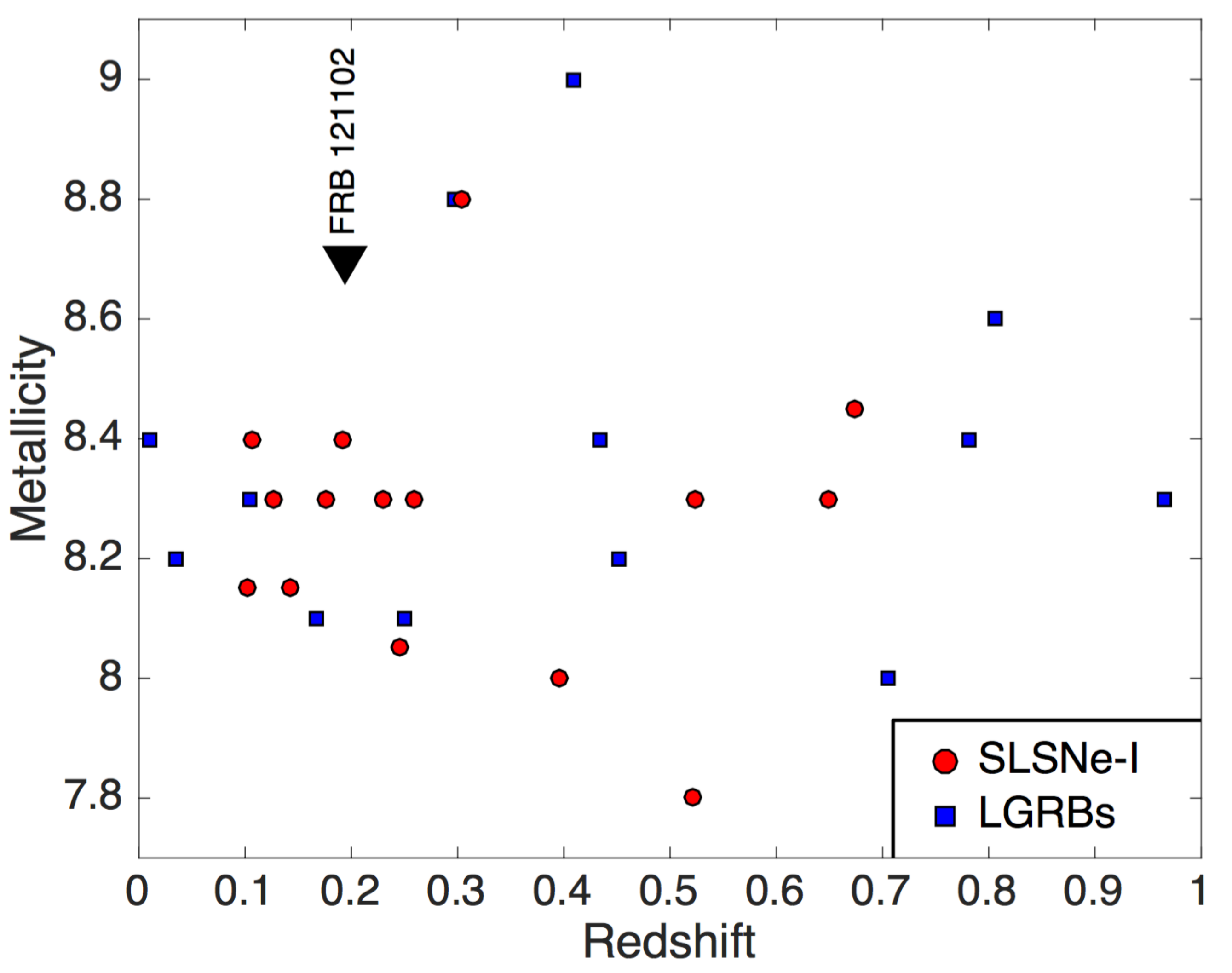}
\caption{Comparison of the absolute magnitude, stellar mass, star formation rate, and metallicity of the host galaxy of \frb to those of SLSNe-I (\citealt{Lunnan+14,Schulze+16,Perley+16}) and LGRBs (\citealt{Modjaz+08,CastroCeron+10,Levesque+10,Vergani+15,Japelj+16,Perley+16d}).}
\label{fig:host}
\end{figure*}

\section{Connection to SLSNe-I and LGRBs}
\label{sec:host}

It is now well established that both LGRBs and SLSN-I exhibit a strong
preference for low-mass, low-metallicity galaxies (e.g.,
\citealt{Stanek+06,Fruchter+06,Modjaz+08,CastroCeron+10,Levesque+10,Lunnan+14,Vergani+15,Schulze+16,Perley+16,Japelj+16,Perley+16d}). \citet{Marcote+17} and
\citet{Tendulkar+17} noted that the host galaxy of \frb\ shares
similar basic properties. Here we provide a
detailed comparison. 

In Figure~\ref{fig:host} compares the
properties of the host of \frb\ to those of LGRBs and SLSN-I at
$z\lesssim 1$. In particular, we show the rest-frame
optical absolute magnitudes, stellar masses, star formation rates, and
metallicities. We find that in all of these fundamental properties,
the host of \frb\ is remarkably similar to the hosts of LGRBs and
SLSN-I. It is located roughly at, or slightly below, the medians of
the LGRB and SLSN-I host galaxy distributions, especially when
considering the low redshift population at $z\lesssim 0.3$.

Beyond their similar host galaxies, additional evidence connects
SLSNe-I and LGRBs to similar stellar progenitors. Both classes tend to
concentrate in bright UV regions of their hosts (e.g.~\citealt{Fruchter+06,Lunnan+14,Schulze+16}). Moreover,
despite clear spectroscopic differences between SLSN-I and the
LGRB-associated SNe near peak brightness (e.g.~\citealt{Liu&Modjaz16}), the nebular spectra (which
probe the explosion ejecta) show close similarity
\citep{Nicholl+16}.  Thus, the differences between LGRB-SNe and
SLSN-I may result from variations in the engine properties, but the
underlying progenitors are likely similar. The remarkable similarity
in the host of \frb\ to those of LGRBs and SLSN-I suggests that it too
shared a similar progenitor and a common engine.

%Given the broadly similar properties of
%their host galaxies of LGRB or a SLSNe-I, it is thus challenging to
%discern from the host of \frb~alone which accompanied it at an earlier
%stage in the system's evolution.

\section{Magnetar Birth in Core Collapse SNe}
\label{sec:magnetar}
We now review the properties of millisecond magnetars at times well after the supernova explosion, when they could be capable of producing a detectable FRB.

\subsection{Magnetars as FRB Sources}

Millisecond period rotation at birth was long predicted to give rise to strong magnetic fields in neutron stars (e.g.~\citealt{Duncan&Thompson92,Thompson&Duncan93}), even prior to the identification of Galactic magnetars \citep{Kouveliotou+98}.  A neutron star with a surface dipole magnetic field strength of $B_{\rm 14} = B_{\rm d}/10^{14}$ G and birth spin period $P_0 = 1P_{\rm ms}$ ms spins down due to magnetic torques, producing a spin-down luminosity\footnote{We have adopted the vacuum dipole spin-down convention employed by Kasen \& Bildsten (2010), which however differs in normalization from the force-free spin-down rate, which is likely more applicable in the plasma-dense environment of a young neutron star (e.g.~\citealt{Spitkovsky06}).}
\begin{eqnarray}
L_{\rm sd} &=& 5\times 10^{46}B_{14}^{2}P_{\rm ms}^{-4}\left(1 + \frac{t}{t_{\rm sd}}\right)^{-2}\,{\rm erg\,s^{-1}} \nonumber \\
&\underset{t \gg t_{\rm sd}}
\approx& 8\times 10^{40}B_{14}^{-2}t_{1}^{-2}\,{\rm erg\,s^{-1}},
\label{eq:Lsd}
\end{eqnarray}
such that its spin period increases with time as
\be
P = P_0\left(1 + \frac{t}{t_{\rm sd}}\right)^{1/2} \underset{t \gg t_{\rm sd}}\approx 28\,{\rm ms}\,B_{14} t_{1}^{1/2},
\label{eq:P}
\ee  
where the spin-down timescale is given by
\be
t_{\rm sd} \simeq 4.7\,{\rm day}\,\,B_{14}^{-2}P_{\rm ms}^{2}.
\label{eq:tsd}
\ee
and we have normalized time $t = 10t_{1}$ yr to a decade following the explosion and in the final equality of eqs.~\ref{eq:Lsd} and \ref{eq:P} we have assumed late-times $t \gg t_{\rm sd}$. 

The rotational energy of a magnetar has been suggested as a power source for both LGRB jets (e.g.~\citealt{Usov92,Wheeler+00,Thompson+04,Bucciantini+08,Metzger+11}) and SLSNe-I (e.g.~\citealt{Kasen&Bildsten10,Woosley10,Metzger+14,Metzger+15a}).  Powering the optical light curve of SLSNe-I requires that the time $t_{\rm sd}$ over which the rotational energy is extracted and deposited behind the ejecta be comparable, or moderately less than, the timescale of the SN optical peak of weeks.  Model fits to SLSNe-I light curves generally find values of $B_{14} \sim 1-10$ and $P \sim 3-5$ ms (e.g.~\citealt{Chatzopoulos+13,Nicholl+14}).  By contrast, powering the relativistic jet of an LGRB typically requires $P \sim 1-2$ ms and much stronger dipole fields $B_{14} \gtrsim 10-30$ in order to match the spin-down luminosity to the beaming-corrected LGRB luminosities of $L_{\gamma} \sim 10^{50}$ erg s$^{-1}$ (e.g.~\citealt{Metzger+11}); however, because of the short timescale of the engine, the light curves of LGRB-SN themselves are powered primarily by radioactive $^{56}$Ni (e.g.~\citealt{Metzger+15a,Cano+16}).\footnote{The ultra-long GRB111209 (engine duration $\approx 1$ hour) was accompanied by an unusually luminous (though not quite super-luminous) SN 2011kl (\citealt{Greiner+15}); this event could represent a hybrid event with
an intermediate-duration engine that produced both a successful jet and later contributed to powering the SN emission (\citealt{Metzger+15a}).} 

We now review ways that a young magnetar could power FRBs through its rotational or magnetic energy, focusing on constraints on the age of the system to explain~\frb.  Each radio burst from \frb~carries an energy $E_{\rm FRB} = f_{\rm b}E_{\rm iso}$, where $E_{\rm iso} \approx 10^{38}-10^{40}$ erg is the measured isotropic equivalent energy (\citealt{Chatterjee+17}) and $f_{\rm b} \lesssim 1$ is the beaming fraction.  Although the value of $f_{\rm b}$ is uncertain, it cannot be too small or the beaming-correct volumetric FRB rate would greatly exceed SLSNe-I/LGRB rates, even when accounting for the fact that a given magnetar magnetar birth event can produce multiple FRBs ($\S\ref{sec:discussion}$).  

Therefore, if the radio bursts are powered by the rotational energy, either directly or in a rotationally powered jet interacting with the stellar ejecta (e.g.~\citealt{Romero+16}), the spin-down luminosity must exceed
\begin{eqnarray}
L_{\rm sd} &\gtrsim& L_{\rm FRB} = \frac{E_{\rm FRB}}{t_{\rm FRB}} \nonumber \\
&\approx& 10^{42}f_{\rm b}\left(\frac{E_{\rm iso}}{10^{39}\rm erg}\right)\left(\frac{t_{\rm FRB}}{\,\rm 1\,ms}\right)^{-1}\,{\rm erg\,s^{-1}},
\end{eqnarray}
where $t_{\rm FRB}$ is the burst duration.  Equation (\ref{eq:Lsd}) shows that, unless $f_{b} \ll 1$, powering an FRB through spin-down requires a very young magnetar of age $t_{\rm age} \lesssim 9\left(L_{\rm FRB}/10^{41}\,{\rm erg\,s^{-1}}\right)^{-1}B_{14}^{-1}$ yr (see also \citealt{Piro16,Lyutikov17}), as well as an extremely efficient mechanism for converting spin-down power into coherent radio emission.  Rotational energy is clearly ruled out for magnetars with ultra-strong dipole fields with $B_{14} \gtrsim 10$ (those capable of powering LGRBs), since the required source age would be shorter than the four years over which \frb~has been observed to burst.

Alternatively, radio bursts may be powered by the release of magnetic energy, similar but potentially more extreme than giant flares produced by Galactic magnetars (\citealt{Giannios10,Popov&Postnov13,Thornton+13,Lyubarsky14}).  \citet{Lyubarsky14} proposed an explicit model for how a magnetic pulse interacting with the magnetar's rotational-powered wind nebula could give rise to an FRB-like burst through synchrotron maser instability.  

Assuming an average internal magnetic field strength of $B_{\rm int} \approx 10^{16}$ G, similar to those expected in the remnants of magnetars at birth (e.g.~\citealt{Moesta+15}), the internal magnetic energy is $E_{\rm B} \approx (4\pi R_{\rm ns}^{3}/3)(B_{\rm int}^{2}/8\pi) \approx 3\times 10^{49}(B_{\rm int}/10^{16} {\rm G})^{2}$ erg, where $R_{\rm ns} = 12$ km is the neutron star radius.  A given magnetar could therefore produce a maximum number of bursts given by
\begin{eqnarray}
N_{\rm FRB} &=& \frac{E_{\rm B}}{E_{\rm FRB}} \nonumber \\
&\approx& 3\times 10^{2}f_{b}^{-1}\left(\frac{f_{\rm r}}{10^{-8}}\right)\left(\frac{B_{\rm int}}{10^{16}\,{\rm G}}\right)^{2}\left(\frac{E_{\rm FRB}}{10^{39}\rm erg}\right)^{-1},
\label{eq:Nburst}
\end{eqnarray}
where $f_{\rm r}$ is the fraction of the flare energy placed into coherent radio emission.  

We have normalized the FRB efficiency to a value of $f_{\rm r} \approx 10^{-8}$, as was estimated by \citet{Lyubarsky14} for a nebula with an electron/positron density of $n_{\pm} \approx 10^{-6}$ cm$^{-3}$.  However, this efficiency scales approximately linearly with the pair density in his model and thus could in principle be much higher given the expected range of pair densities $n_{\pm} \sim 10^{-7}-10^{-3}$ cm$^{-3}$ in the young magnetar nebulae on timescales of decades (depending on the pair multiplicity of the wind; see eq.~\ref{eq:npair}).  On the other hand, searches for a coincident radio counterpart to the giant flare from the Galactic magnetar SGR 1806-20 ruled out radio emission similar to the observed bursts from \frb~\citep{Tendulkar+16}, potentially implying a lower radiative efficiency.  

\citet{Spitler+14} observed 11 bursts from \frb~in about 0.6 days of total observing time with an average isotropic energy per burst of $\approx 4\times 10^{38}$ erg; extended over the $\gtrsim$ 4 year period of burst activity, this suggests a total number of bursts of $\gtrsim 2\times 10^{4}$ bursts for $f_b \sim 1$, already comparable or exceeding the estimate of $N_{\rm FRB}$ from equation (\ref{eq:Nburst}) for the expected range of $f_{\rm r}$.  Energetic constraints may therefore be pointing to a system lifetime not greatly exceeding the current bursting duration, although this again depends on the uncertain value of $f_b$.

If the FRB source were in fact giant magnetic flares, we would expect coincident gamma-ray emission from \frb, of luminosity $L_{\gamma} \sim 10^{47}$ erg s$^{-1}$ similar to Giant flares in our galaxy (e.g.~\citealt{Palmer+05}).  However, this emission would be too dim to detect at the distance of \frb~by {\it Fermi} GBM or {\it Swift} BAT.  Even if the young magnetar produced quiescent X-ray emission above the {\it Chandra} upper limits of $\sim 5\times 10^{41}$ erg s$^{-1}$, the oxygen-rich SN ejecta remains opaque to photoelectric absorption at keV energies for decades or longer after the explosion (Appendix, eq.~\ref{eq:tauX}).  

%In summary, in either spin-down or magnetic-powered models for the radio bursts, the magnetar is unlikely to be more than a few decades to a century old.

\subsection{Transparency of the SN Ejecta}
\label{sec:ejecta}

\begin{figure}
\includegraphics[width=.5\textwidth]{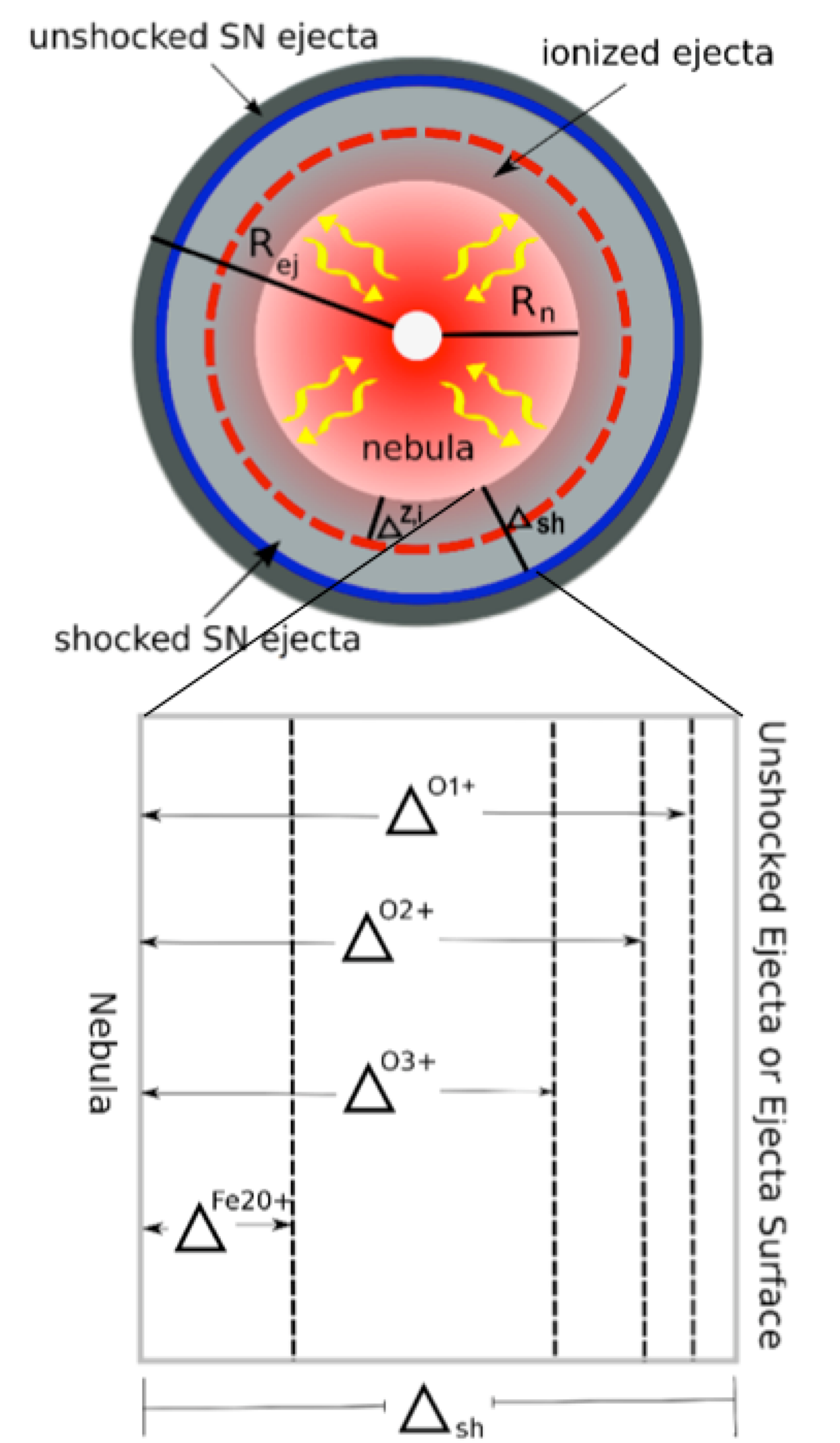}
\caption{Schematic illustration of the magnetar wind nebula embedded in the expanding SN ejecta (adapted from a similar figure in \citealt{Metzger+14}).  UV/X-ray radiation from the nebula ionizes through the ejecta from within, similar to an HII region (see Appendix).  }
\label{fig:cartoon}
\end{figure}

Additional constraints on the age of the system result from requiring that the GHz radio emission can escape through the expanding SN ejecta, as well as the requirement to not overproduce the maximum local contribution to the DM or its derivative (see also, e.g., \citealt{Connor+16,Piro16}).  

We consider the stellar progenitor to be a compact, stripped-envelope star, similar to those of LGRBs and SLSNe-I.  We adopt characteristic values of $M_{\rm ej} \approx 10M_{\odot}$ and $v_{\rm ej} \approx 10^{4}$ km s$^{-1}$ for the mass and mean velocity of the ejecta, motivated by observations of SN associated with LGRBs and SLSNe-I (e.g., \citealt{Nicholl+15}).  These fiducial values imply a kinetic energy of $\approx M_{\rm ej} v_{\rm ej}^{2}/2 \approx 10^{52}$ erg, similar to those of hyper-energetic LGRB-SN and some SLSNe-I.

The free expansion phase of the SN blast wave lasts until it sweeps up a mass comparable to its own in the pre-explosion stellar wind or interstellar medium.  The density profile of a steady wind is given by $\rho_{\rm w} = \dot{M}_{\rm w}/(4\pi r^{2}v_{\rm w}) = A/r^{2}$, where $\dot{M}_{\rm w}$ is the wind mass-loss rate and $v_{\rm w} \approx 1000$ km s$^{-1}$ is the wind velocity of the compact progenitor star.  We normalize the wind parameter $A = \dot{M}_{\rm w}/(4\pi v_{\rm w})$ according to the standard convention (e.g.~\citealt{Chevalier&Li99}),
\be
A_{\star} \equiv \frac{A}{5\times 10^{11}\rm \,g\,cm^{-1}\,} \approx 1.0\left(\frac{\dot{M}_w}{10^{-5}M_{\odot}\,{\rm yr^{-1}}}\right)\left(\frac{10^{3}\,{\rm km\,s^{-1}}}{v_{\rm w}}\right)
\ee
and thus consider values $A_{\star} \sim 0.1-10$, given the uncertain mass loss history prior to the explosion.  The majority of the ejecta will slow down once it reaches the deceleration radius $R_{\rm dec} = M_{\rm ej}/(4\pi A)$ at which its swept up equals its own; this occurs on a timescale of\footnote{In reality, the steady wind profile will become invalid well before $R_{\rm dec}$, but this doesn't affect our subsequent conclusions.}
\be
t_{\rm dec} = \frac{R_{\rm dec}}{v_{\rm ej}} = \frac{M_{\rm ej}}{4\pi A v_{\rm ej}} \approx 10^{5}{\rm yr}\,\,A_{\star}^{-1}M_{1}v_{9}^{-1},
\label{eq:tST}
\ee
where $M_{1} \equiv M_{\rm ej}/10M_{\odot}$ and $v_{\rm 9} \equiv v_{\rm ej}/10^{4}$ km s$^{-1}$.  Thus, at times $t \ll t_{\rm dec}$ the mean ejecta radius increases as 
\be R_{\rm ej} = v_{\rm ej}t \approx 0.11 \,{\rm pc}\,v_{9}t_{1}
\ee and hence will remain smaller than the 5GHz VLBI upper limit on the quiescent radio source \frb~of $\lesssim 0.7\,{\rm pc}$ \citep{Marcote+17} for $t \lesssim 70$ yr for typical parameters.

On timescales exceeding substantial rotational energy input from the magnetar ($t \gg t_{\rm sd}$; eq.~\ref{eq:tsd}), the ejecta will approach homologous expansion with a mean density decreasing approximately as $\rho_{\rm ej} = (3M_{\rm ej}/4\pi R_{\rm ej}^{3}) \propto t^{-3}$.  The corresponding mean free electron density is 
\be
n_{\rm e} \simeq \frac{3M_{\rm ej}f_{\rm ion}}{8\pi R_{\rm ej}^{3}m_p} = 4.5\times 10^{4}\,{\rm cm^{-3}}\,f_{\rm ion}M_{1}v_{9}^{-3}t_{1}^{-3},
\ee
where $f_{\rm ion}$ is the ionized fraction of the ejecta, which in general will vary with radius out through the ejecta (see below).  Based on the anticipated nucleosynthesis in engine-driven SNe (e.g.~\citealt{Maeda+02}), we assume for simplicity that the ejecta is composed entirely of oxygen, though in detail it will contain smaller mass fractions of other elements with mass to charge ratio $A/Z = 2$ like helium, carbon, and iron.  

The resulting plasma frequency within the ejecta is
\be
\nu_{\rm p} = \frac{1}{2\pi}\sqrt{\frac{4\pi  n_{\rm e}e^{2}}{m_e}} \simeq 0.002\,{\rm GHz}\,f_{\rm ion}^{1/2}M_{1}^{1/2}v_{9}^{-3/2}t_{1}^{-3/2},
\ee
and hence, even for $f_{\rm ion} \approx 1$, is no significant barrier to the escape of GHz radiation on timescales of years after the explosion.

A more stringent constraint comes from the free-free optical depth of the ejecta for our assumed oxygen-dominated composition.  Absent external radiation, the expanding SN ejecta will cool adiabatically, becoming almost neutral ($f_{\rm ion} \ll 1$) once the temperature drops below a few thousand K on a timescale of months after the explosion.  However, the ejecta eventually becomes re-ionized by two processes.  First, a reverse shock is produced as the SN ejecta decelerates upon colliding with the progenitor wind, with a high enough temperature to completely reionize the outer layers of the ejecta (\citealt{Piro16}).  A second potential source of ionization, which we focus on, comes from the UV/X-ray radiation of the central rotationally-powered magnetar nebula, which acts to reionize the ejecta over time from within (\citealt{Metzger+14}).  This produces multiple ionization fronts, such that $f_{\rm ion}$ decreases with radius from the inner edge of the ejecta directly exposed to nebula out to the ejecta surface, as illustrated in Fig.~\ref{fig:cartoon}. 

If the nebula produces comparable energy across a range of photon frequencies from soft UV to soft X-rays, then in general low ionization states (e.g.~OI, OII) with threshold ionization frequencies in the UV are easier to photo-ionize than higher states (e.g.~OVII, OVIII) with ionization frequencies at soft X-rays.  In the Appendix we estimate that the nebula may completely ionize OI and OII on a timescale of a decade or less, depending sensitively on the ejecta mass and the magnetic field of the magnetar.  However, complete ionization (up to OVIII) probably requires centuries or longer.  On timescales of a few decades to a century, the radius- or mass-averaged ionized fraction is thus expected to range from $f_{\rm ion} \approx 0$ (if the ejecta remains entirely neutral) to $f_{\rm ion} \approx 2/Z = 0.25$ if OI$-$OII are ionized.

The free-free optical depth through the ejecta shell for an oxygen-dominated composition ($Z = 8, A = 16$) is 
\begin{eqnarray}
\tau_{\rm ff} &=& (0.018 Z^{2}\nu^{-2}T_{\rm ej}^{-3/2}n_{\rm e}n_{\rm ion}\bar{g}_{\rm ff})R_{\rm ej} \nonumber \\
 &\approx& 93 \bar{g}_{\rm ff} f_{\rm ion}^{2}\nu_{\rm GHz}^{-2}T_{4}^{-3/2}M_{1}^{2}t_{1}^{-5}v_{9}^{-5}
\label{eq:tauff}
\end{eqnarray}
where $n_{\rm ion} \approx 2n_{\rm e}/A$ is the ion density, $T_{\rm ej} = 10^{4}T_{4}$ K is the ejecta temperature normalized to a typical value for photo-ionized gas and $\bar{g}_{\rm ff} \sim 1$ is the Gaunt factor. 

 The ejecta becomes transparent to free-free absorption ($\tau_{\rm ff} < 1$) after a time 
\be
t_{\rm ff} \simeq 9.85\,{\rm yr} \,\,\bar{g}_{\rm ff}^{1/5} \left(\frac{f_{\rm ion}}{0.1}\right)^{2/5}\nu_{\rm GHz}^{-2/5}T_{4}^{-3/10}M_{1}^{2/5}v_{9}^{-1}
\ee
of order decades for fiducial parameters.  

A even more stringent constraint comes from the DM through the ejecta shell, 
\be
{\rm DM}_{\rm ej} \simeq n_{\rm e}R_{\rm ej} \approx 462\,{\rm pc\,cm^{-3}}\,\left(\frac{f_{\rm ion}}{0.1}\right)M_{1}v_{9}^{-2}t_{1}^{-2},
\ee
which at a minimum must be less than the local contribution of DM$_{\rm local} < $ DM$_{\rm host + local} \approx 55-225$ pc cm$^{-3}$ (\citealt{Tendulkar+17}).  Another related constraint comes from the time derivative of the dispersion measure,
\be
\left|\frac{d{\rm DM}_{\rm ej}}{dt}\right| = \frac{2{\rm DM}_{\rm ej}}{t} \approx 92\,{\rm pc\,cm^{-3}\,yr^{-1}}\,\left(\frac{f_{\rm ion}}{0.1}\right)M_{1}v_{9}^{-2}t_{1}^{-3},
\label{eq:dDMdt}
\ee
which \cite{Piro16} estimate obeys $d{\rm DM}_{\rm ej}/dt \lesssim 2$ pc cm$^{-3}$ yr$^{-1}$ for \frb.  

If just OI is ionized, then $f_{\rm ion} \approx 0.1$ and both constraints result in a similar minimum source age of $t \gtrsim 30$ yr for fiducial parameters.  By contrast, if OI remains neutral because the UV radiation field of the nebula is weak, then the age could be younger, while if OII is ionized the minimum age could approach a century.    

Using our oxygen ionization model developed in the Appendix, Figure \ref{fig:DM} compares the time evolution of DM and $d$DM/$dt$ for different assumptions about the magnetar field strength, SN ejecta mass, and the fraction of the luminosity of the magnetar wind nebula in UV ionizing photons, $\epsilon_{\rm ion}$.  Whether the source age constraint is closer to 10 or 100 years old requires a more detailed model for the ionization structure of the ejecta, including a more accurate model for the SED of the magnetar wind nebula.

%By combining equations (\ref{eq:tauff}) and (\ref{eq:dDMdt}) we can also write the DM derivative directly in terms of the free-free optical depth as
%\begin{eqnarray}
%\left|\frac{d{\rm DM}_{\rm ej}}{dt}\right| &\approx& 99\,{\rm pc\,cm^{-3}\,yr^{-1}}\times \nonumber \\
%&\approx&\,\tau_{\rm ff}^{3/5}\bar{g}_{\rm ff}^{-3/5} \left(\frac{f_{\rm ion}}{0.1}\right)^{-1/5}M_{1}^{-1/5}v_{9}T_{4}^{9/10}\nu_{\rm GHz}^{6/5},
%\label{eq:dDMdt2}
%\end{eqnarray}
%such that $d{\rm DM}_{\rm ej}/dt \lesssim 2$ pc cm$^{-3}$ yr$^{-1}$ for $\tau_{\rm ff} \lesssim 10^{-3}$, in which case we would be observing \frb~within a timescale of a few decades after it became optically thin at GHz frequencies.  Figure \ref{fig:DM} compares the time evolution of DM and $d$DM/$dt$ for different assumptions about the magnetar field strength, SN ejecta mass, and the fraction of the luminosity of the magnetar wind nebula in UV ionizing photons, $\epsilon_{\rm ion}$.

%In summary, both free-free ($\tau_{\rm ff} < 1$) and DM constraints can also be overcome after timescales of a few decades to a century for typical properties of LGRB SN or SLSNe-I, depending sensitively on the ionized fraction.  Whether the source is closer to being 10 or 100 years old requires a more detailed model for the ionization structure of the ejecta, including a realistic SED for the magnetar wind nebula.

\begin{figure*}
\includegraphics[width=0.5\textwidth]{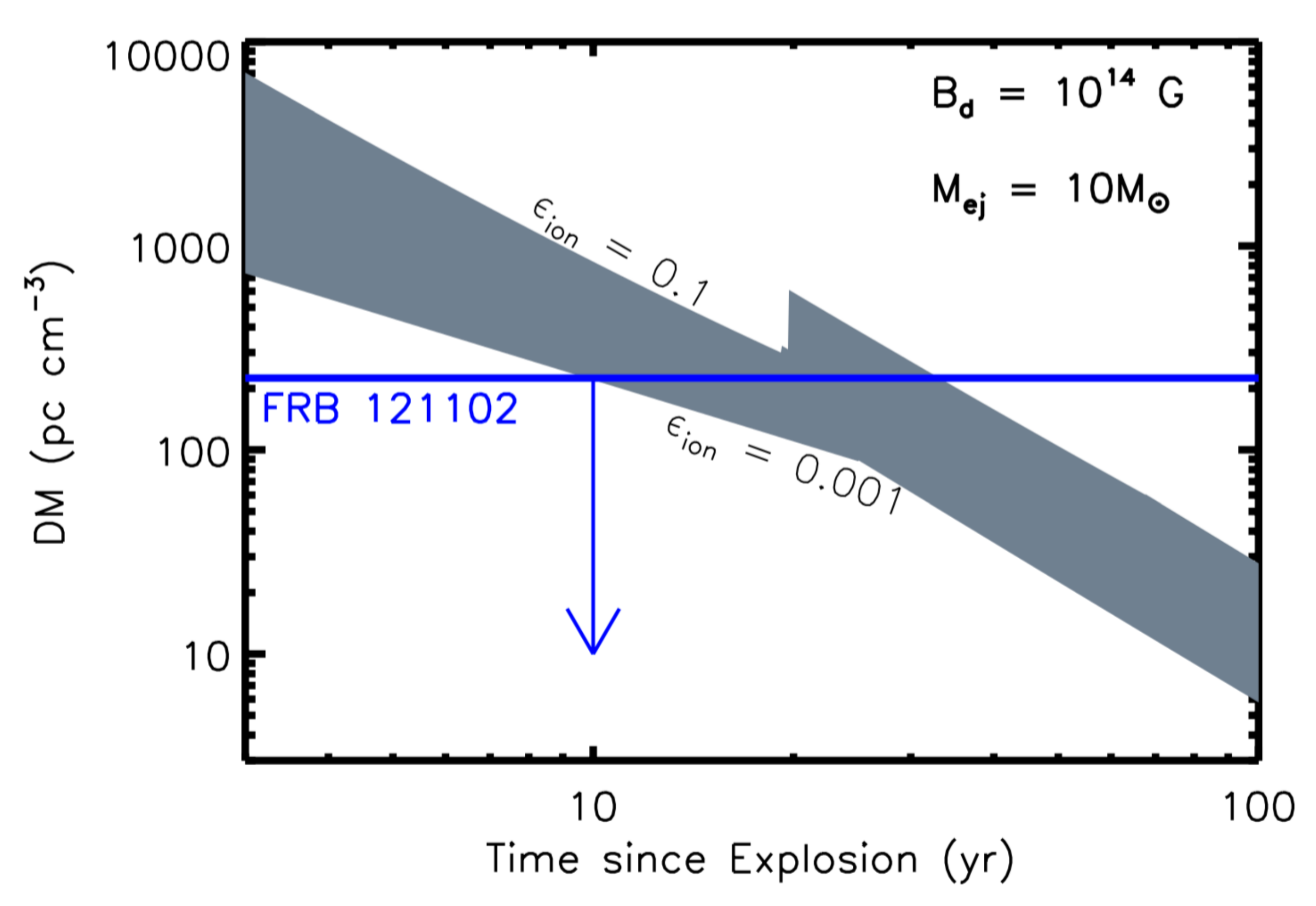}
\includegraphics[width=0.5\textwidth]{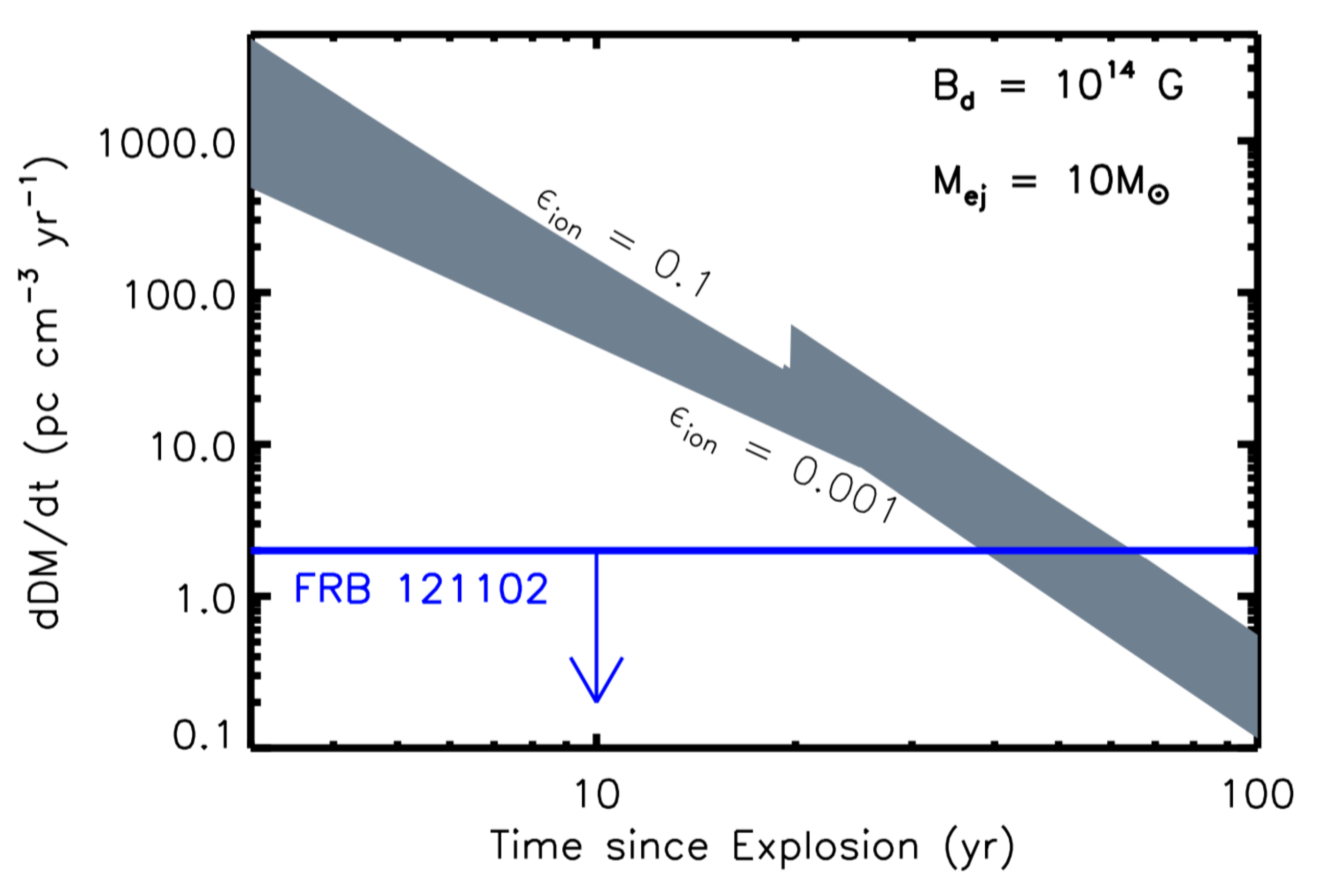}
\includegraphics[width=0.5\textwidth]{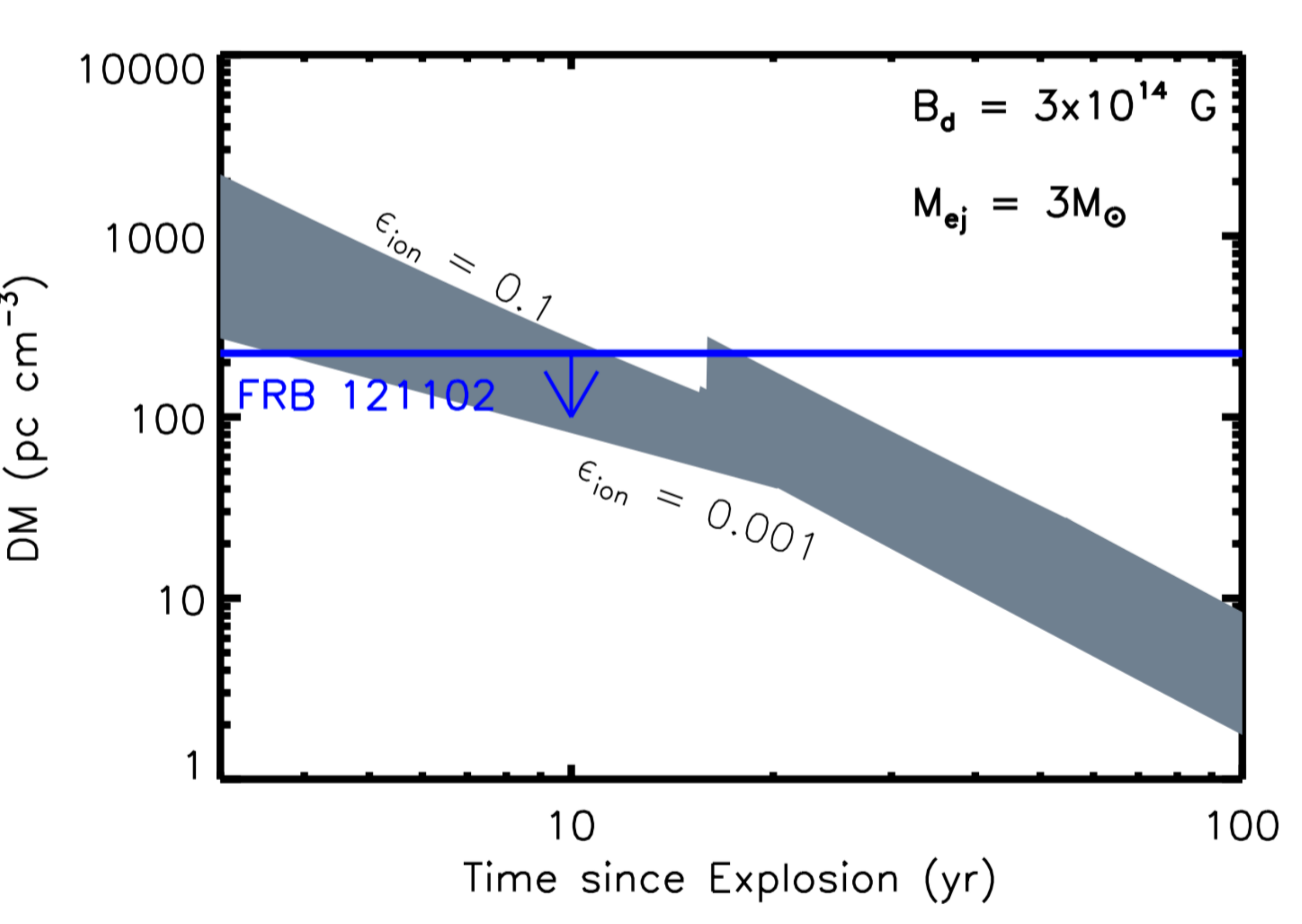}
\includegraphics[width=0.5\textwidth]{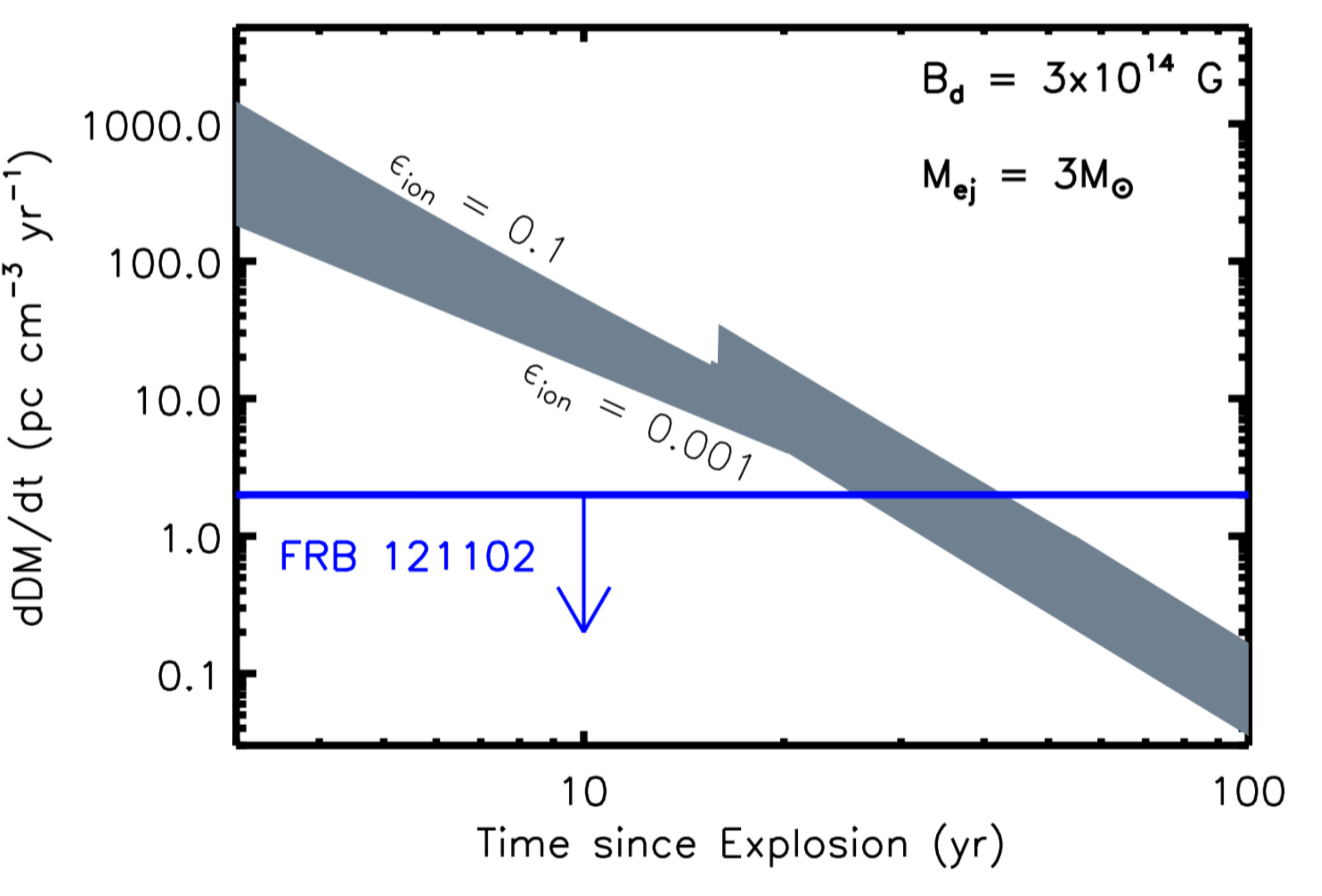}
\caption{The grey region shows the dispersion measure (DM) (left panels) and its time derivative (right panels) through the expanding oxygen-rich SN ejecta as a function of time since the explosion across a range of values for the fraction $\epsilon_{\rm ion} = 0.001-0.1$ of the magnetar luminosity placed into ionizing UV radiation.  We have assumed an ejecta velocity of $v_{\rm ej} = 10^{9}$ cm s$^{-1}$ and electron temperature in the ionized layer of $T = 10^{4}$ K (see Appendix).  The top and bottom panels are shown for different values of the magnetar dipole field strength $B_{\rm d} = 10^{14}, 3\times 10^{14}$ G and total SN ejecta mass $M_{\rm ej} = 3,10M_{\odot}$, respectively.  }
\label{fig:DM}
\end{figure*}

%is also consistent with {\it Chandra} upper limits on the 0.5-10 keV luminosity of $L_{\rm X} \lesssim 6\times 10^{41}\,{\rm erg\,s^{-1}}$ (\citealt{Chatterjee+16}).  If oxygen is not fully ionized, the photo-electric optical depth of the ejecta is given by
%\be
%\tau_{\rm X} \simeq \frac{\rho_{\rm ej}\sigma_{\rm bf}}{A m_p} \approx 130 M_{1}v_{9}^{-2}t_{1}^{-2}\left(\frac{E_X}{1\,{\rm keV}}\right)^{-3}
%\ee

\section{Source of Quiescent Radio Emission}
\label{sec:radio}

We consider three possible sources for the quiescent synchrotron radio source, each reasonably expected on decade to century timescales following the birth of a millisecond magnetar.  These include (1) emission from the rotationally-powered magnetar wind nebula ($\S\ref{sec:nebula}$); (2) external shock interaction between the magnetar-energized supernova blast wave and the wind of the progenitor star ($\S\ref{sec:shock}$); (3) an orphan afterglow from an initially off-axis LGRB that accompanied the core collapse event ($\S\ref{sec:orphan}$).  

We first point out a general constraint on the size of the quiescent radio-emitting region, $R_{\rm rad}$, which is set by the brightness temperature,
\be
T_{\rm B} = \frac{c^{2}D^{2}F_{\nu}}{2\pi kR_{\rm rad}^{2}\nu^{2}} \approx 2.5\times 10^{9}{\rm K}\left(\frac{F_{\nu}}{250 \mu\rm Jy}\right)\nu_{\rm GHz}^{-2}\left(\frac{R_{\rm rad}}{1\,{\rm pc}}\right)^{-2}
\label{eq:TB}
\ee
The observed radio spectrum shows a flat spectral index $\beta \approx -0.2$ at frequencies $\gtrsim$ 1.6 GHz (\citealt{Chatterjee+17}).   However, synchrotron self-absorption should instead produce a rising spectrum with $\beta \approx 2-2.5$ for brightness temperatures exceeding the `temperature' of  relativistic electrons of Lorentz factor $\gamma_e \sim 2-100$ contributing to the radio emission in this range, e.g.~$T_{\rm B} \gtrsim \gamma_e m_e c^{2}/k \approx 6\times 10^{10}(\gamma_e/10)$ K.  The observed 1.6 GHz flux of $F_{\nu} \approx 250\mu$Jy therefore requires an emitting size 
\be R_{\rm rad} \gtrsim 0.13 (\gamma_e/10)^{-1/2}\,{\rm pc}. \label{eq:Rrad}\ee
While this is consistent with the 5GHz VLBI size constraint $R_{\rm rad} \lesssim 0.7$ pc (\citealt{Marcote+17}), it suggests that the quiescent radio source - whatever its nature - should possess a self-absorption break not too far below the GHz band.  In fact, the observed flattening of the spectral index moving to lower frequencies may be part of a gradual SSA transition (Fig.~\ref{fig:spectrum}).  

Also note that if the radio bursts must pass through the emitting region of the quiescent source in escaping the environment of the burst (as is the case in all of our proposed scenarios) then the low frequency spectrum of the FRBs should also experience significant synchrotron absorption  (see also \citealt{Yang+16}), possibly  contributing to the non-detections of FRBs to date at low radio frequencies  (e.g.~\citealt{Karastergiou+15,Caleb+16,Rowlinson+16}).

\subsection{Magnetar Wind Nebula}
\label{sec:nebula}

As in other young pulsar wind nebulae (\citealt{Gaensler&Slane06}), the rotationally powered wind from the magnetar inflates a nearly spherical bubble of relativistic electron/positron pairs behind the SN ejecta (\citealt{Metzger+14,Murase+16}).  The characteristic radius of the nebula is given by \citep{Chevalier77}
\be
R_{\rm n} \approx 1.2L_{\rm sd}^{1/5} M_{\rm ej}^{-1/5}v_{\rm ej}^{3/5}t^{6/5} \approx 0.03{\rm pc} \,B_{14}^{-2/5}M_{1}^{-1/5}v_{9}^{3/5}t_{1}^{4/5},
\ee
which will remain smaller than quiescent source upper limit of 0.7 pc for a couple centuries.  The quiescent radio source could therefore represent synchrotron radiation from this nebula, since if the ejecta is transparent to free-free absorption to a GHz frequency burst it will also be transparent at higher frequencies.  A flat radio spectrum is also a common observational feature of PWNe; the radio spectral index $\beta \approx -0.25$ of the Crab Nebula (\citealt{Bietenholz+97}) is remarkably similar to the low frequency bands of the quiescent source associated with \frb.    

We can estimate the magnetic field of the nebula at time $t$ by assuming that the magnetic energy  $(B_{\rm n}^{2}/8\pi)V_{\rm n}$, where $V_{\rm n} \approx 4\pi R_{\rm n}^{3}/3$ is the nebula volume, is a fraction $\epsilon_B$ of the injected spin-down energy $\sim L_{\rm sd}t$,
\be
B_{\rm n} \simeq \left(\frac{6\epsilon_{B} L_{\rm sd}t}{R_{\rm n}^{3}}\right)^{1/2} \simeq 0.04{\rm G}\,\epsilon_{B,-2}^{1/2}t_{1}^{-17/10} B_{14}^{-2/5}M_{1}^{3/10}v_{9}^{-9/10},
\ee
where we have normalized $\epsilon_{B} = 10^{-2}\epsilon_{B,-2}$ to a value similar to the magnetization inferred for the Crab Nebula (e.g.~\citealt{Kennel&Coroniti84}).

There are several locations within the nebula where electrons could be accelerated to relativistic velocities.  
One source of particle acceleration is the wind termination shock (\citealt{Kennel&Coroniti84}).  Electron/positron pairs are carried out by the pulsar wind at a rate 
$\dot{N}_{\pm} = \mu_{\pm}\dot{N}_{\rm GJ}$, where $\dot{N}_{\rm GJ} = 8\pi^{2}B_{\rm d}P^{-2}R_{\rm NS}^{3}/ec$ is the Goldreich-Julian flux and $\mu_{\pm}$ is the pair multiplicity.  The number density of freshly-injected\footnote{Since $\dot{N}_{\rm GJ} \propto t^{-1}$ at times $t \gg t_{\rm sd}$, the number of pairs injected to the nebula is approximately the same per logarithmic time interval.} in the nebula is thus approximately
\be
n_{\pm} \simeq \frac{\dot{N}_{\pm}t}{V_{\rm n}} \approx 1.3\times 10^{-5}\,{\rm cm^{-3}}\left(\frac{\mu_{\pm}}{10^{2}}\right)B_{14}^{1/5}M_{1}^{3/5}v_{9}^{-9/5}t_{1}^{-7/5},
\label{eq:npair}
\ee
where we have normalized the pair multiplicity to value $\mu_{\pm} \approx 10^{2}$ expected for young magnetar winds (e.g., \citealt{Medin&Lai10,Beloborodov13}).  

If these pairs are accelerated impulsively at the wind termination shock carrying most of the total spin-down luminosity, they will reach a random pair Lorentz   
\be
\gamma_{\pm} \simeq \frac{L_{\rm sd}}{\dot{N}_{\pm} m_e c^{2}} \approx 7.5\times 10^{8}\left(\frac{\mu_{\pm}}{10^{2}}\right)^{-1}B_{14}^{-1}t_{1}^{-1}
\ee
However, the synchrotron frequency of such pairs,
\begin{eqnarray}
h\nu_{\rm m} &=& \frac{h}{2\pi}\frac{eB_{\rm n}\gamma_{\pm}^{2}}{m_e c} \nonumber \\
 &\approx& 260\,{\rm MeV}\,\left(\frac{\mu_{\pm}}{10^{2}}\right)^{-2} M_{1}^{3/10}B_{14}^{-12/5}v_{9}^{-9/10}t_{1}^{-3.7}
\end{eqnarray}
is typically in the X-ray to gamma-ray frequency range on timescales of decades to a century (depending on $\mu_{\pm}$), much too high to explain the observed GHz radio emission, unless the pair multiplicity is orders of magnitude higher than in well-observed PWNe.  Furthermore, extending a synchrotron spectrum $\nu F_{\nu} \propto \nu^{4/3}$ from the UV/X-ray to radio band produces a spectral index inconsistent with that measured from \frb~and a flux which is much too low to explain the measured values.

However, radio emission from the Crab Nebula does not conform to the simplest picture in which all particle acceleration occurs due to diffusive shock acceleration at the pulsar wind termination shock.  The observed radio frequency spectral index of $\beta \approx -0.25$ (\citealt{Bietenholz+01}) of optically thin synchrotron emission (across three decades in frequency) requires an energy distribution of the emitting electrons $dN/dE \propto E^{-p}$ with slope $p = 2\beta + 1 \approx 1.5$.  This ``excess'' of low energy electrons has been attributed to pairs ejected at a much earlier phase, when the spin-down power was higher, whose energy has been degraded by adiabatic expansion (\citealt{Atoyan99}).  Such an excess could be produced in a very young magnetar remnant due to the much higher pair creation rate due to $\gamma-\gamma$ annihilation, as is expected less than months after the explosion while the compactness parameter of the nebula is still $\gg 1$, leading to a pair formation cascade (\citealt{Metzger+14,Metzger&Piro14}).  More detailed modeling is needed to determine whether the energy in currently slow-cooling relic pairs can explain the quiescent source.

On the other hand, observations of time variable `wisps' in the radio band of the Crab Nebula (\citealt{Bietenholz+01}) show that at least some radio-emitting electrons are being accelerated currently, in the same region as the higher energy emission.  This has led to the suggestion of other acceleration sites than the termination shock, such as the magnetic reconnection in the striped pulsar wind (e.g.~\citealt{Sironi&Spitkovsky11,Zrake&Arons16}), which particle-in-cell plasma simulations show can indeed produce flatter ($p < 2$) electron spectra (e.g., \citealt{Sironi&Spitkovsky15}), consistent with that needed to explain a flatter spectral index $\beta \approx -0.2$.  

We consider that such an anomalous particle distribution $p = 1.5$ is also at work in young magnetar nebulae, and that it carries a majority of the pulsar power.  For such a particle distribution $p < 2$, most of the total  energy is in high energy electrons.  

Electrons with random Lorentz factors $\gamma$ above a critical value
\be \gamma_c = \frac{6\pi m_e c}{B_{\rm n}^{2}\sigma_T t} \approx 1500\epsilon_{B,-2}^{-1}B_{14}^{4/5}v_9^{9/5}M_{1}^{-3/5}t_{1}^{12/5} \ee
will have a synchrotron cooling timescale shorter than the ejecta expansion timescale $t$.  The synchrotron cooling frequency is thus given by
\be
\nu_c = \frac{1}{2\pi}\frac{eB_{\rm n}\gamma_c^{2}}{m_e c} \approx 250\,{\rm GHz}\,\epsilon_{B,-2}^{-3/2}B_{14}^{6/5}v_9^{27/10}M_{1}^{-9/10}t_{1}^{3.1}
\ee

Therefore, if the particle distribution extends to a maximum Lorentz factor $\gamma_{\rm max} \gtrsim \gamma_{\rm c}$, with a corresponding maximum synchrotron frequency $\nu_{\rm max} \gtrsim \nu_{\rm c}$, the radio flux in the GHz band can be written as 
\begin{eqnarray}
 F_{\nu} &\approx& \frac{L_{\rm sd}}{8\pi D^{2}\nu_{\rm max}}\left(\frac{\nu}{\nu_{\rm max}}\right)^{-0.25} \nonumber \\
&\approx& 200\,{\rm \mu Jy}\nu_{\rm GHz}^{-0.25}\left(\frac{\nu_{\rm max}}{10^{12}\,{\rm Hz}}\right)^{-0.75}B_{14}^{-2}t_{1}^{-2}
\label{eq:Fnu}
\end{eqnarray}
Thus, for $\nu_{\rm max} \lesssim 10^{12}$ Hz and $B_{14} \sim 1$ we can reach $\sim 1-10$ GHz radio fluxes comparable to the values $\approx 200 \mu$Jy measured from the quiescent counterparts of \frb~on timescales of decades.  For comparison, in the Crab Nebula the value of $\nu_{\rm m}$ (occurring near the peak of the SED) occurs at $\nu_{\rm max} \approx 10^{14}$ Hz.  The present day radio emission could also be boosted from equation (\ref{eq:Fnu}) by slow-cooling pairs ejected at earlier times (years or less after the explosion), when the magnetar wind power $L_{\rm sd} \propto t^{-2}$ was substantially higher.

Equation \ref{eq:Fnu} also makes clear that the nebula scenario is not a viable source of radio emission on decade timescales for magnetars with ultra-strong fields ($B_{14} \gtrsim 10$) hypothesized to power LGRB jets, again because they spin down too quickly.

\subsection{External Blast Wave}
\label{sec:shock}

\begin{figure}
\includegraphics[width=0.5\textwidth]{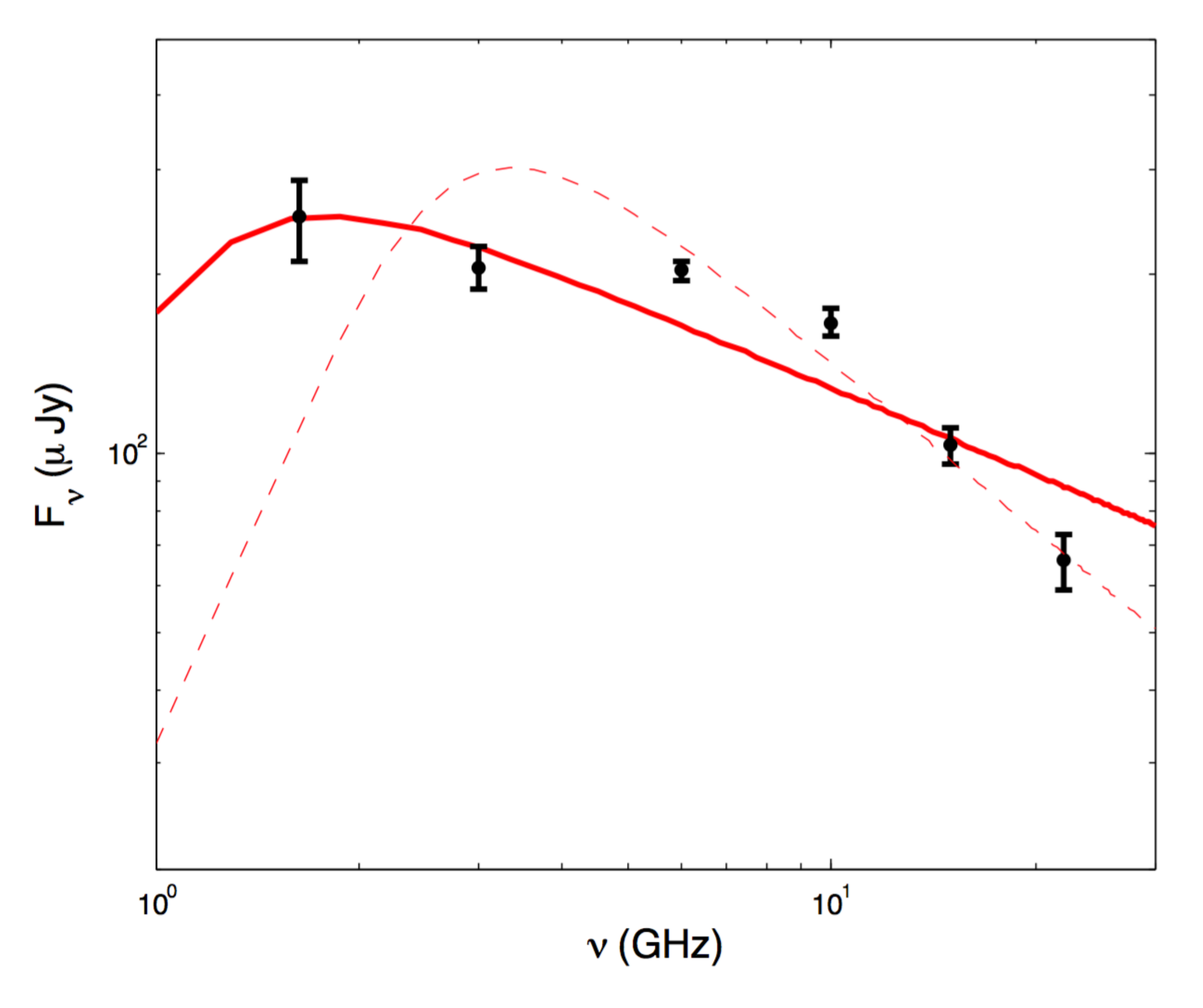}
\caption{Spectrum of the quiescent radio source coincident with \frb~\citep{Chatterjee+17}, compared to a model for a self-absorbed synchrotron spectrum of the form $F_{\nu} = F_0\left(\nu/\nu_a\right)^{5/2}\left(1-\exp\left[-(\nu/\nu_a)^{-(p+4)/2}\right]\right)$ \citep{Chevalier98}, where $\nu_a$ is the self-absorption frequency and $p$ is the electron acceleration index.  A solid line shows the best-fit model ($F_0 \approx 350\mu$Jy, $\nu_a = 1.2$ GHz, $p = 1.967$), while a dashed line shows a ``high $p$" model for which $F_0 = 460 \mu$Jy, $\nu_a = 2.9$ GHz, $p = 2.9$.  Although neither fit is satisfactory, the precise frequency dependence of the SSA roll over is sensitive to the geometry of the aborbing layer.  }
\label{fig:spectrum}
\end{figure}

Another source of synchrotron radiation is shock interaction between the fastest layers of the SN ejecta and the surrounding wind of the progenitor star.  The nebula inflated by the magnetar drives a shock through the ejecta, which reaches the surface of the star on a timescale comparable to the spin-down time $\sim t_{\rm sd}$ if the total rotational energy from the magnetar exceeds the initial kinetic energy of the ejecta of $\approx 10^{51}$ erg  (e.g.~\citealt{Kasen+16,Chen+16}), i.e. for magnetar birth spin periods of $P \lesssim 2-3$ ms.  

Two-dimensional hydrodynamical simulations of this process of bubble inflation by \citet{Suzuki&Maeda16} show that final density distribution of the matter reaches a homologous state with the distribution of density with velocity given by $\rho(v) \propto v^{-\beta}$, where $\beta \approx 5-6$ 
in the outer ejecta, and $\beta \approx 0$ in an inner core of the ejecta
(hereafter we adopt $\beta = 6$ in the outer ejecta, and a flat $\beta = 0$ inner core). 
The transition velocity between outer-inner ejecta profiles $v_{\rm t}$ is therefore related to the total ejecta mass and energy by $v_{\rm t} = \sqrt{10 E_{\rm ej} / 3 M_{\rm ej}}$.
The ejecta energy distribution above $v_{\rm t}$ obeys
$E_{>v} = (3 M_{\rm ej}v_{\rm t}^{2}/4)(v/v_{\rm ej})^{-1}$,
and the adiabatic-shock dynamics are governed by energy conservation, \citep[e.g.][]{Margalit&Piran15}
\be
E_{>v} \approx M_{\rm dec}(R_{\rm dec}) v^2 .
\ee 
This implies a deceleration radius {\it for a given mass layer} to sweep up its own energy in the progenitor stellar wind
\be
R_{\rm dec}(v) = \frac{3 M_{\rm ej}}{16 \pi A} \left(\frac{v}{v_{\rm t}}\right)^{-3} .
\ee
Modifying eq.~(\ref{eq:tST}), we find that the deceleration time for a mass layer at velocity $v$ is therefore
\be
t_{\rm dec}(v) = \frac{3 R_{\rm dec}}{4 v}
 \approx 5.7 \times 10^{4}{\rm yr}\,\,A_{\star}^{-1}M_{1}(v/v_{\rm ej})^{-4}v_{9}^{-1},
\ee
where the prefactor $3/4$ in the first equality results from solving the forward-shock dynamic equation.
The layer undergoing deceleration at a given time ($t = t_{\rm ST}$) has velocity
\be
v_{\rm dec} = 8.7\times 10^{9}\,{\rm cm\,s^{-1}}\,\,\, t_{1}^{-1/4}A_{\star}^{-1/4}M_{1}^{1/4}v_{9}^{3/4},
\ee
mass
\begin{eqnarray}
M_{\rm dec} = 4\pi A R_{\rm dec} 
%\nonumber \\
\approx 0.012
M_{\odot}\,\,\, t_{1}^{3/4}A_{\star}^{3/4}M_{1}^{1/4}v_{9}^{3/4}
\end{eqnarray}
and kinetic energy
\be
E_{\rm dec} = E_{>v_{\rm dec}} \approx 1.7 \times 10^{51}\,\,{\rm erg}\,\, t_{1}^{1/4}A_{\star}^{1/4}M_{1}^{3/4}v_{9}^{9/4}
\ee
Deceleration of this layer occurs on a radial scale
\be
R_{\rm dec} = \frac{4}{3} v_{\rm dec}t \approx 3.7\times 10^{18}{\rm cm}\,\,t_{1}^{3/4}A_{\star}^{-1/4}M_{1}^{1/4}v_{9}^{3/4},
\label{eq:rdec}
\ee
again marginally consistent with the upper limits on the 5 GHz source size \citet{Marcote+17} and the lack of self-absorption at $\nu \gtrsim 1.6$ GHz (eq.~\ref{eq:Rrad}) for timescales of decades and $A_{\star} \sim$ 10 (as needed to explain the observed radio fluxes; see below).

The shocked gas will accelerate relativistic electrons, with a characteristic Lorentz factor
\be
\gamma_m = \epsilon_e \frac{m_p}{m_e}\frac{p-2}{p-1}\left(\frac{v_{\rm dec}}{c}\right)^{2} \underset{p = 2.3}\approx 3.3\epsilon_{e,-1}t_{1}^{-1/2}A_{\star}^{-1/2}M_{1}^{1/2}v_{9}^{3/2},
\ee
where we have made the standard assumption of a shock-accelerated electron distribution $dN/dE \propto E^{-p}$ with $p \approx 2.3$ (characteristic of other trans-relativistic shocks) that carries a total fraction $\epsilon_e = 0.1\epsilon_{e,-1}$ of the shock power.  The shock-generated magnetic field, distributed throughout the shocked volume $\approx 4\pi R_{\rm dec}^{3}/3$, can be estimated as
\be
B = \left(\frac{6\epsilon_B E_{\rm dec}}{R_{\rm dec}^{3}}\right)^{1/2} \approx 4.6\times 10^{-3}{\rm G}\, \epsilon_{B,-1}^{1/2}t_{1}^{-1}A_{\star}^{1/2},
\ee
where again $\epsilon_{B} = 10^{-1}\epsilon_{B,-1}$ is the equipartition fraction.  The resulting characteristic synchrotron frequency is
\be
\nu_{m} = \frac{1}{2\pi}\frac{eB\gamma_m^{2}}{m_e c} \approx 1.61\times 10^{5}{\rm Hz}\,\,\epsilon_{e,-1}^{2}\epsilon_{B,-1}^{1/2} t_{1}^{-2}A_{\star}^{-1/2}M_{1}v_{9}^{3}
\label{eq:num}
\ee
while the flux density at the distance of \frb\, ($D = 3\times 10^{27}$ cm) is given by
\be
F_{\nu_m} = \frac{N_e}{4\pi D^{2}}\frac{m_e c^{2}\sigma_T}{3e} B \approx 9.9{\rm mJy}\,\,\epsilon_{B,-1}^{1/2}t_{1}^{-1/4}A_{\star}^{5/4}M_{1}^{1/4}v_{9}^{3/4}
\ee
where $N_e = M_{\rm dec}/(2 m_p)$ is the number of shocked electrons.
The flux at higher frequencies $\nu > \nu_{\rm m}$ is given by
\begin{eqnarray}
F_{\nu} &=& F_{\nu_m}\left(\frac{\nu}{\nu_{m}}\right)^{-(p-1)/2} \nonumber \\
&\underset{p = 2.3}\approx& 34\mu{\rm Jy} \,\, \nu_{\rm GHz}^{-0.65}\epsilon_{B,-1}^{0.83}\epsilon_{e,-1}^{1.3}t_{1}^{-1.55}A_{\star}^{0.925}M_{1}^{0.9}v_{9}^{2.7}, \nonumber \\
\label{eq:Fnublast}
\end{eqnarray}
where in the second line we have taken $p = 2.3$.  We thus see that its possible to reproduced the observed fluxes of $F_{\nu} \approx 200 \mu$Jy in the GHz frequency range for $A_{\star} \sim $ 10 and $\epsilon_e \approx 0.2$. 

Finally, the synchrotron self-absorption frequency can be expressed as \citep{Chevalier98} 
\begin{eqnarray}
\nu_{\rm a} &\approx& 0.1{\rm GHz} \,\, \xi(p) \epsilon_e^{\frac{2p-2}{p+4}} \epsilon_B^{\frac{p+2}{2(p+4)}} A_{\star}^{\frac{5-p}{2(p+4)}} M_{\rm ej}^{\frac{2p-3}{2(p+4)}} v_{\rm t}^{\frac{6p-9}{2(p+4)}} t^{-\frac{4p+5}{2(p+4)}}
\nonumber \\  &\underset{p=2.3}\approx& 0.14{\rm GHz} \,\, \epsilon_{e,-1}^{0.41} \epsilon_{B,-1}^{0.34} A_{\star}^{0.21} M_{1}^{0.13} v_9^{0.38} t_{1}^{-1.13}
\label{eq:nuSA}
\end{eqnarray}
where $\xi(p)$ is an order unity prefactor. For $\epsilon_{e} \approx 0.2$ and $A_{\star} \sim 10$ we therefore expect the self-absorption frequency to fall marginally below the observational band, $\nu_m < \nu_{\rm a} \lesssim 1{\rm GHz}$.

One concern with this scenario is that the observed spectral index of $\beta \approx -0.2$ is flatter than the value $\beta = -0.65$ assumed here.  If interpreted as optically-thin synchrotron emission $\beta \approx -0.2$ would appear to require $p \approx 2\beta + 1 \approx 1.4$, an electron distribution  which as already discussed is extremely challenging to produce from particle acceleration at a shock.   However, as pointed out at the beginning of this section, synchrotron self-absorption could be setting in just below the frequency window of observations (eq.~\ref{eq:TB} and surrounding discussion).  This implies that the true asymptotic spectral index in the frequency range of a few GHz could be steeper than the inferred value of $\beta \approx -0.2$; indeed,  at the highest radio frequencies $\nu \gtrsim 10$ GHz there is evidence for a steeper index closer to $\beta \approx -1$.  

To illustrate this point, Figure \ref{fig:spectrum} shows the spectrum of the quiescent radio source coincident with \frb~\citep{Chatterjee+17}, compared to a model for a self-absorbed synchrotron spectrum of the form $F_{\nu} = F_0\left(\nu/\nu_a\right)^{5/2}\left(1-\exp\left[-(\nu/\nu_a)^{-(p+4)/2}\right]\right)$ from \citet{Chevalier98}, where $\nu_a$ is the self-absorption frequency and $p$ is the electron acceleration index.  A solid line shows the best-fit model ($F_0 \approx 350\mu$Jy, $\nu_a = 1.2$ GHz, $p = 1.97$), while a dashed line shows a ``high $p$" model for which $F_0 = 460\mu$Jy, $\nu_a = 2.9$ GHz, $p = 2.9$.  Although these fits are not altogether convincing, given the sensitive dependence on the spectral roll-off on the precise structure of the absorbing layer, it seems possible that self absorption could explain the observed spectral hardening at low frequencies.  

Another concern with this model is that radio emission has not yet been detected in coincidence with SLSNe-I, even in cases where relatively tight constraints are available, such as the nearby SLSNe-I SN 2015bn ($z = 0.1$) for which upper limits on the 7.4(22) GHz radio emission on a timescale of about 1 year were in the range $F_{\nu} \lesssim 40(75) \mu$Jy (\citealt{Nicholl+16b}), corresponding to limits roughly 4 times deeper at the larger distance of \frb.  However, if the self-absorption frequency is indeed just below the 1 GHz band on timescales of a few decades for \frb, then on timescales of a year or less the SA frequency $\nu_{\rm sa} \propto t^{-1.1}$ (eq.~\ref{eq:nuSA}) would be higher and hence the radio emission might be suppressed from the value predicted by equation \ref{eq:Fnublast}.    

%Despite these issues, we conclude that circumstellar shock interaction remains a viable contender for the quiescent radio source.  

%One additional check on this scenario is confirm that the free-free optical depth of the shocked wind material is insufficient to block the FRB from escaping.  Recasting equation (\ref{eq:tauff}) for the external shocked wind ($f_{\rm ion} \approx 1$) we find
%\begin{eqnarray}
%\tau_{\rm ff} &=& (0.07\nu^{-2}T_{\rm dec}^{-3/2}n_{\rm dec}^{2}\bar{g}_{\rm ff})R_{\rm dec} \approx 2\times 10^{-8} \bar{g}_{\rm ff}\nu_{\rm GHz}^{-2}A_{\star}^{y}M_{1}^{x}t_{1}^{y}v_{9}^{z}
%\end{eqnarray}
%where $n_{\rm dec} = 3M_{\rm dec}/(8\pi m_p R_{\rm dec}^{3}) \approx 0.05 t_{1}^{-1}v_{9}^{-1}$ cm$^{-3}$ and we take $T_{\rm dec} \approx 3m_p v_{\rm dec}^{2}/16k_b \approx 2\times 10^{11}t_{1}^{-2/3}A_{\star}^{-1/3}M_{1}^{2/3}v_{9}^{4/3}$ K as the temperature of the shocked gas.

\subsection{Radio Afterglow from an Off-Axis LGRB Jet}
\label{sec:orphan}

\begin{figure}
\includegraphics[width=.5\textwidth]{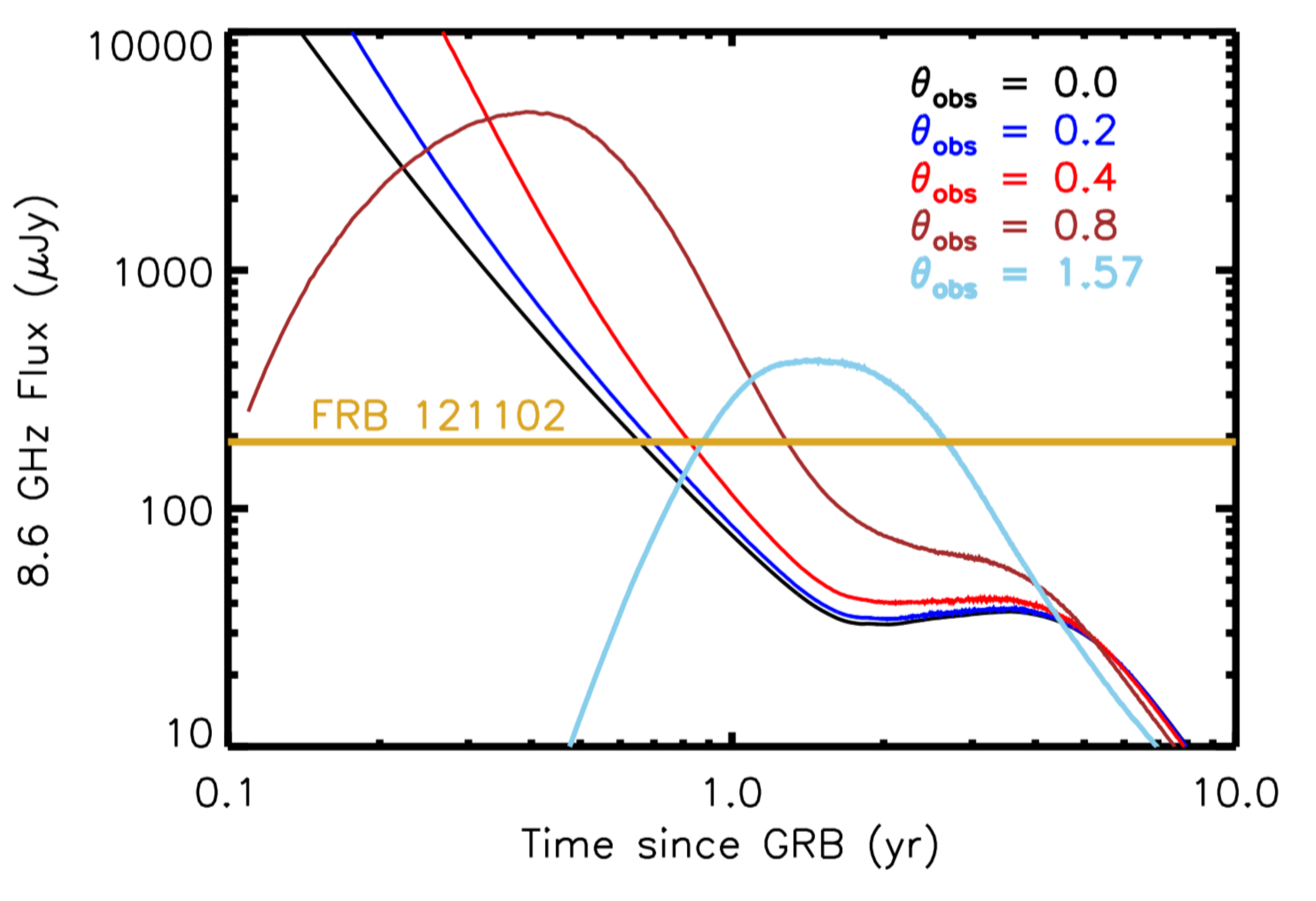}
\caption{Model for off-axis orphan radio afterglow emission of a GRB jet for different viewing angles relative to the jet axis \citep{VanEerten+10} compared to the flux density of~\frb.  The jet kinetic energy is $E_{\rm k} = 2\times 10^{51}$ erg and propagates into an external medium of constant density $n = 1$ cm$^{-3}$.  The parameters of the shock-accelerated electrons are $p = 2.5$, $\epsilon_e = \epsilon_B = 0.1$. }
\label{fig:afterglow}
\end{figure}

Millisecond proto-magnetars, along with black hole accretion, are contenders for the central engines of LGRBs (e.g.~\citealt{Thompson+04}).  LGRB jets produce ultra-relativistic ejections in tightly collimated jets with opening angles $\theta_{j} \ll 1$ (\citealt{Frail+01}).   Assuming that the radio bursts from \frb~are isotropic, or at least not directed along the GRB jet, then the GRB jet would most likely be pointed away from our line of sight.  However, as material slows down by shocking the interstellar medium, even off-axis viewers enter the causal emission region of the synchrotron afterglow (\citealt{Rhoads97}).  When viewed in an initial off-axis direction, the emission from such an `orphan afterglow' can at late times become approximately isotropic once the shocked matter decelerates to sub-relativistic velocities (e.g.~\citealt{Zhang&MacFadyen09,Wygoda+11}).  

Figure \ref{fig:afterglow} compares the 8.6 GHz off-axis LGRB afterglow models of \citet{VanEerten+10} for different viewing angles $\theta_{\rm obs}$ relative to the axis of the jet to the current flux of the quiescent radio source associated with \frb~(gold line).  The predicted off-axis jet emission drops below the emission from \frb~for all viewing angles within roughly one year for moderately on-axis models $\theta_{\rm ob} \lesssim 0.8$ and on a timescale of $\sim$ 3 years for the completely off-axis model $\theta_{\rm obs} = 1.57$.  This would appear to disfavor the orphan afterglow explanation for the quiescent radio counterpart, since \frb~has been undergoing bursts for at least $\sim 4$ years, placing an absolute lower limit on the time since the jet was launched.  However, the peak flux and peak time of the  afterglow depend on several uncertain parameters (jet energy, density and radial structure of the external medium, electron acceleration efficiency $\epsilon_e$, etc.), such that in principle the radio flux could stay brighter for longer.  The 4.9 GHz flux of LGRB 030329 (at a redshift $z = 0.168$ similar to\frb; \citealt{Stanek+03,Hjorth+03}) at $t \approx 10$ yr after the burst was $F_{\nu} \approx 20\mu$Jy (\citealt{Mesler+12}), somewhat brighter than predicted by the models in Fig.~\ref{fig:afterglow} on a similar timescale, and is decaying approximately as $F_{\nu} \propto t^{-1}$.

\section{Discussion}
\label{sec:discussion}

Most of our proposed  explanations for the quiescent radio source require a relatively young remnant, at most a few decades in age (or possibly less in the case of an off-axis LGRB).  This can indeed be comparable to the timescale over which the oxygen-rich ejecta is becoming transparent to GHz radio, and the dispersion measure and its derivative decreasing to values consistent with observations ($\S\ref{sec:ejecta}$), although this depends sensitively on the degree to which the magnetar is able to ionize the ejecta (Appendix).  Due to selection effects associated with the decaying rotational and magnetic energy of the magnetar likely causing the luminosity of bursts to also decay with time, we might expect that detected FRB sources will be dominated by those produced just as the ejecta is becoming optically thin to radiation.  This also implies that a negative time derivative of the DM could well be measured soon (eq.~\ref{eq:dDMdt})

Such a scenario, in which we preferentially observe FRBs almost as soon it becomes possible, is consistent with the lack of FRB discovery at sub-GHz frequencies (\citealt{Karastergiou+15,Caleb+16,Rowlinson+16}), since free-free absorption (e.g.~\citealt{Piro16}) or synchrotron absorption from the same emission region responsible for the quiescent radio source (eq.~\ref{eq:TB} and surrounding discussion) become more severe at lower frequencies (Fig.~\ref{fig:spectrum}).  Propagation of the pulse through the magnetar wind nebula might also induce excess Faraday rotation from that expected due to propagation through the Galaxy and IGM, as  measured by \citet{Masui+15} for FRB 110523 \citep{Piro16}.  
%Furthermore, the long scattering tails observed in some FRB are also inconsistent with all the DM originating from the IGM (\citealt{Luan&Goldreich14}).

A potential obstacle to a claimed association between FRBs and SLSNe-I or LGRBs are their relative rates.  
\citet{Howell+13} calculate the volumetric rate of SLSNe-I at $z \sim 1$ to be $91^{+76}_{-36}$ yr$^{-1}$ Gpc$^{-3}$ (see also \citealt{McCrum+15}).  This is comparable to the estimated beaming-correct local ($z \approx 0$) rate of LGRBs of $\approx 130^{+60}_{-70}$ Gpc$^{-3}$ yr$^{-1}$ for an assumed beaming fraction of $f_{\rm b} = 1/100$ (\citealt{Wanderman&Piran10}).  These are substantially lower, by a factor of $\sim 10-100$ than the estimate volumetric FRB rate, using their observed rate of a $\lesssim 10^{4}$ per sky per day of bursts $\gtrsim 1$ Jy and assuming distances derived assuming most of the excess DM is extragalactic (\citealt{Thornton+13,Law+15,Rane+16,Scholz+16}; Vander Wiel et al.~2016). 

 However, given the observation of that some FRBs repeat (producing multiple bursts per object), the LGRB/SLSNe-I rate could well be compatible with the birth rate of FRB-producing objects.  In such a scenario, most FRBs are repeating, but we only observe the very brightest ones, with \frb~being an exception due to the higher sensitivity of Areceibo \citep{Spitler+14}.  Furthermore, somewhat fine-tuned parameters are required in the magnetar model to give an extremely luminous supernovae (e.g.~\citealt{Kasen&Bildsten10}) or a LGRB \citep{Metzger+11}, implying that the true population of millisecond magnetar-forming SNe could be higher than estimated from these populations.         

We now describe possible tests of the proposed SLSNe-I/LGRB/FRB association.  On several decade timescales after the explosion, the magnetar spin period will increase from an initial value of a few ms to $P \approx 30-300$ ms (eq.~\ref{eq:P}).  Although no periodicities have been observed in the repeated bursts of \frb~previously, this provides motivation to continue search for periodicities in this range.  We might also expect to detect FRBs or their quiescent radio counterparts from the locations of previous LGRBs or SLSNe-I.\footnote{There have been previous suggestions connecting FRBs and LGRB (\citealt{Zhang14,Deng&Zhang14,Gao+14}), but most of these are describing time coincidences.  Time coincidence searches between FRBs and LGRBs (\citealt{Yamasaki+16}) led to the claimed discovery of a gamma-ray counterpart to FRB 131104 by \citet{DeLaunay+16}.  If this association is confirmed, there must be multiple classes of FRBs, including both `catastrophic' and 'repeating' types.  
}  Given that accurate localizations have only been available for a large number of GRBs since the launch of {\it Swift} in 2005, and that the first SLSNe-I were discovered in 2005 and 2006 (SN2005ap;  \citealt{Quimby+07}, and SCP06F6; \citealt{Barbary+09}), the ability to monitor LGRB/SLSNe-I on $\gtrsim$ decade timescales is only now becoming possible.  The known distance to the source could also aid such searches by reducing the possible range of DM.  Higher frequency observations are preferred due to free-free absorption, especially in the case of searches for analogs to the quiescent source from \frb.  

Finally, in most of our scenarios we expect the quiescent radio source in \frb~to fade on a timescale comparable to the inferred system age of decades or a century, such that 10\% decay could be visible within a few years.  Furthermore, at some point relatively soon a negative time derivative of the burst DM should be measured, though a more detailed model of the ionization evolution of the ejecta is required to better quantify this.  Given the complex multidimensional structure of the interaction between the magnetar nebula and the ejecta and its susceptibility to hydrodynamic instabilities (e.g.~Rayleigh-Taylor fingers; \citealt{Chen+16,Suzuki&Maeda16}) a complex, non-monotonic time evolution of the DM along a given line of sight might well be expected.

\section{Conclusions}
\label{sec:conclusions}

We propose that the association between the host galaxy \frb~and the properties of the hosts of SLSNe-I and LGRBs \citep{Chatterjee+17,Tendulkar+17} can be understood if both classes of objects are associated with the birth of a millisecond magnetar.  While the LGRB and SLSNe-I occur within minutes and weeks, respectively, of the core collapse, the escape of an FRB from the supernova ejecta instead requires a timescale of a decade or longer.  This timescale, which we argue is controlled mainly by photo-ionization by the magnetar wind nebula (Appendix), could occur somewhat earlier than in normal core collapse SNe.  This is in part because the ejecta velocities of magnetar-powered hypernovae can be $\sim 3$ times higher than those of normal core collapse events and in part due to the fact that oxygen-rich ejecta is more challenging to ionize than hydrogen-rich ejecta.  

Following the submission of this paper, \citet{Kashiyama&Murase17} proposed an association of \frb~with a young pulsar embedded in the low mass ejecta of an ``ultra-stripped" hydrogen-poor supernova  (e.g.~\citealt{Kleiser&Kasen14,Tauris+15}), a model that has been previously invoked to explain the observed population of fast-evolving Type Ic SNe (e.g.~\citealt{Drout+14}).  However, we note that the host galaxy properties of the ultra-fast stripped SNe, such as their location in the mass-metallicity plane consistent with that comprising the bulk of star-forming galaxies, do not match those of the hosts of \frb~or LGRBs/SLSNe-I (\citealt{Drout+13,Drout+14}), thus disfavoring such an association.

We have shown several ways in which a continuum radio source observed within a parsec of the location of \frb~would be naturally expected in the magnetar scenario, provided again that the source age is indeed at most a few decades to a century old.  These include emission from the magnetar wind nebula (in analogy with other PWNe like the Crab Nebula) or synchrotron emission from the fastest supernova ejecta or from an off-axis LGRB afterglow.  Distinguishing between these possibilities could be aided by further low frequency observations of the bursts or quiescent source, to look for evidence of free-free absorption and to ascertain whether both emission sources are passing through the same ejecta shell.  Constraints on the flux and source size already imply that synchrotron self-absorption should becoming relevant at energies just below the current observing band.     

 Importantly, all of our proposed explanations for the quiescent radio source connect to an {\it engine-powered explosion}.  An initially rapidly rotating (millisecond) magnetar or high-field pulsar is an inevitable feature the model to explain the quiescent radio source, even if the FRB pulses themselves are more likely powered by the dissipation of magnetic energy.  Millisecond magnetars, perhaps produced preferentially in metal-poor environments like those characterizing the host of \frb, could conceivably be distinct in terms of their giant flare properties from the Galactic population of magnetars, which are mostly produced at solar metallicity or above.

Finally, we have suggested possible tests of the claimed association between FRBs an SLSNe-I by monitoring the locations of $\gtrsim$ decade old LGRBs and SLSN-I with arcsecond positions for coherent radio bursts and/or analogs to the quiescent radio source seen in association with \frb.       Measurement of the time derivative of the DM would tightly constrain the system age and is expected for a young source.   Finally, the quiescent source should fade appreciably over the coming years to a decade.  

\acknowledgements

We thank Dimitrios Giannios and Maxim Lyutikov for helpful comments.  BDM gratefully acknowledges support from the National Science Foundation (AST-1410950, AST-1615084), NASA through the Astrophysics Theory Program (NNX16AB30G) and the Fermi Guest Investigator Program (NNX15AU77G, NNX16AR73G), the Research Corporation for Science Advancement Scialog Program (RCSA 23810), and the Alfred P.~Sloan Foundation.

\clearpage

\appendix

\section{Photo-Ionization of Supernova Ejecta by Magnetar Nebula}

In this section we estimate the conditions for the magnetar nebula to photo-ionize various atomic species through the ejecta shell as a function of time after the explosion, in order to assess the free-free opacity of the ejecta shell to radio emission and the photo-electric opacity to soft X-rays.  For simplicity we consider that the ejecta is composed entirely of oxygen (\citealt{Maeda+02}).  

A given atomic species $i$ will be ionized to a radial depth $\Delta_{\rm ion}$ through the ejecta shell where the bound-free\footnote{We neglect scattering opacity relative to absorptive opacity since at late times of interest the ejecta is optically thin to Thomson scattering.} optical depth of a photon near the ionization threshold energy $h\nu_{\rm th}$ 
reaches unity (\citealt{Metzger+14}; see Fig.~\ref{fig:cartoon}), 
\be
\Delta_{\rm ion} \simeq \frac{2}{\rho_{\rm ej}\kappa_{\rm bf}(\nu_{\rm th})} = \frac{2A m_p}{\rho_{\rm ej}\sigma_{\rm bf}f_{\rm n}},
\label{eq:deltaion}
\ee
where $\kappa_{\rm bf} = (f_{\rm n}\sigma_{\rm bf}/Am_p$) is the bound-free opacity and
\be
\sigma_{\rm bf,\nu} \simeq \sigma_{\rm th}\left(\frac{\nu}{\nu_{\rm th}}\right)^{-3}, \nu \gtrsim \nu_{\rm th}
\ee
is the bound-free cross section and $\sigma_{\rm th}$ is the cross section at threshold.  The factor of 2 in equation (\ref{eq:deltaion}) results because we have estimated the penetration depth as twice that of a photon with the threshold frequency $\nu = \nu_{\rm thr}$ (see Appendix B of \citealt{Metzger+14}).

We assume that a fraction $\epsilon_{\rm ion} \sim 0.001-0.1$ of the magnetar spin-down power $L_{\rm sd}$ (eq.~\ref{eq:Lsd}) is placed into the radiation energy distribution of the nebula $E_{\nu}$ near the threshold ionization frequency $\nu_{\rm th}$, i.e. the ionization radiation energy density of the nebula incident on the ejecta is
\be
\nu_{\rm th} u_{\rm ion,\nu} = \epsilon_{\rm ion}\frac{L_{\rm sd}}{4\pi R_{\rm ej}^{2}c}
\ee

The ``neutral'' (non-ionized) fraction $f_{\rm n}$ of species $i$ in the layer directly exposed to the nebula is determined by the competition between the rates of photo-ionization and radiative recombination according to 
\begin{eqnarray}
f_{\rm n} &\simeq& \left(1 + \frac{4\pi}{\alpha_{\rm rec}n_{\rm e}}\int \frac{J_{\nu}}{h\nu}\sigma_{\rm bf,\nu}d\nu\right)^{-1} \underset{f_{n} \ll 1}\approx \frac{\alpha_{\rm rec}M_{\rm ej}f_{\rm ion} h\nu_{\rm th}}{m_p \epsilon_{\rm ion}L_{\rm sd}t v_{\rm ej}\sigma_{\rm th}} \nonumber \\
&\approx& 7.6\times 10^{-4}\left(\frac{0.1}{\epsilon_{\rm ion}}\right)f_{\rm ion}M_{1}B_{14}^{2}t_{1}\left(\frac{h\nu_{\rm th}}{10\,{\rm eV}}\right)\frac{\alpha_{\rm rec,-11}}{\sigma_{\rm th,-18}},
\label{eq:fn}
\end{eqnarray}
where $\sigma_{\rm th,-18} = \sigma_{\rm th}/(10^{-18}$ cm$^{2}$), $\alpha_{\rm rec} = 10^{-11}\alpha_{-11}$ cm$^{3}$ s$^{-1}$ is the rate of radiative recombination, $J_{\nu_{\rm th}} = c u_{\rm ion,\nu}/4\pi$ is the mean intensity of the nebula near the ionization threshold incident on the ejecta of volume $V_{\rm ej} \simeq 4\pi R_{\rm ej}^{3}/3$.

On decade timescales of interest, equation (\ref{eq:fn}) in most cases we expect that $f_{\rm n} \ll 1$, in which case the thickness of the photo-ionized layer (eq.~\ref{eq:deltaion}) can be estimated as
\begin{eqnarray}
\frac{\Delta_{\rm ion}}{R_{\rm ej}} \simeq \frac{8\pi Am_p^{2} v_{\rm ej} \epsilon_{\rm ion}L_{\rm sd} tR_{\rm ej}^{2}}{3 f_{\rm ion} M_{\rm ej}^{2}\alpha_{\rm rec}(h\nu_{\rm th})} 
\approx 1.4\left(\frac{\epsilon_{\rm ion}}{0.1}\right)M_{1}^{-2}f_{\rm ion}^{-1}\alpha_{-11}^{-1}v_{9}^{3}t_{1}B_{14}^{-2}\left(\frac{h\nu_{\rm th}}{10\,{\rm eV}}\right)^{-1}
\end{eqnarray}
A given species will thus become completely ionized throughout the ejecta once $\Delta_{\rm ion} \approx R_{\rm ej}$, as occurs after a time
\be
t_{\rm ion} \approx 7.1\,{\rm yr}\left(\frac{\epsilon_{\rm ion}}{0.1}\right)^{-1}M_{1}^{2}f_{\rm ion}\alpha_{-11}v_{9}^{-3}B_{14}^{2}\left(\frac{h\nu_{\rm th}}{10\,{\rm eV}}\right)
\ee
Moving from OI to OVIII, threshold energies increase $h\nu_{\rm th} =$ [13.6, 35, 55, 77, 110, 140, 740, 870] eV; the threshold cross sections are $\sigma_{\rm th,-18} = $ [12,9,4,1.6,0.8,0.3,0.2,0.09] (\citealt{Verner+96}); and the recombination rates at $T \approx 10^{4}-3\times 10^{4}$ K generally increase $\alpha_{-11} = $ [0.004-0.04, 0.04-0.09, 1.3-1.0, 4.0-3.0, 0.5-1.5, 0.9-0.3, 3.6-1.7, 4.6-2.2] (\citealt{Nahar&Pradhan97}).  Finally, the ionization fraction penetrating the $i$th ionization layer will increase as $f_{\rm ion} = i/8$.  

The net effect of the increasing values of $h\nu_{\rm th}$, $\alpha_{\rm rec}$, and $f_{\rm ion} = i/8$ is that, although it is possible to ionize OI and OII within a few years for a characteristic value of $\epsilon_{\rm ion} \sim 0.01$, $t_{\rm ion}$ can easily become much longer than decades for higher ionization states.  Note also the sensitive dependence on $t_{\rm ion}$ with the magnetic field, such that for $B_{14} = 3$ even OI will remain neutral for decades ($f_{\rm ion} \approx 0$) for $\epsilon_{\rm ion} = 0.01$.  We thus conclude that on timescale of decades $f_{\rm ion}$ will vary from $\approx 0$ to $\approx 2/Z \sim 0.25$. 

Because it is very unlikely that 10$M_{\odot}$ of oxygen will be completely ionized on timescales of interest, X-ray photons of energy $\sim 1-10$ keV $\gtrsim h\nu_{\rm th} = 0.87$ keV will be strongly attenuated by photo-electric absorption.  The X-ray optical depth is given by
\be
\tau_{X} \approx   \frac{\rho_{\rm ej}\sigma_{\rm bf}R_{\rm ej}}{A m_p} \approx 100 \left(\frac{E_{X}}{1\,\rm keV}\right)^{-3}t_{1}^{-2}v_{9}^{-2},
\label{eq:tauX}
\ee
such that the ejecta will remain opaque to X-rays of energies $E_{\rm X} \lesssim $ few keV for a decade or longer.

\bibliographystyle{yahapj}
\bibliography{ms}
\end{document}